\documentclass[preprint]{aastex}

\newcommand{\mylabel}[1]{\label{#1}}
\newcommand{\myref}[1]{\ref{#1}}

\begin{document}

\title{An Adaptive Grid, Implicit Code for Spherically Symmetric,
General Relativistic Hydrodynamics in Comoving Coordinates}

\author{Matthias Liebend\"{o}rfer\altaffilmark{1,2,3},
Stephan Rosswog\altaffilmark{1,4},
and Friedrich-Karl Thielemann\altaffilmark{1,2}
}
\altaffiltext{1}{Department of Physics and Astronomy, University of Basel,
Klingelbergstrasse 82, 4056 Basel, Switzerland}
\altaffiltext{2}{Physics Division, Oak Ridge National Laboratory, Oak Ridge,
Tennessee 37831-6354}
\altaffiltext{3}{Department of Physics and Astronomy, University of Tennessee,
Knoxville, Tennessee 37996-1200}
\altaffiltext{4}{Department of Physics and Astronomy, University of Leicester,
LE1 7RH Leicester, United Kingdom}


\begin{abstract}
We describe an implicit general relativistic hydrodynamics code. The
evolution equations are formulated in comoving coordinates. A
conservative finite differencing of the Einstein equations is outlined
and artificial viscosity and numerical diffusion are discussed. The
time integration is performed with AGILE, an implicit solver for stiff
algebro-differential equations on a dynamical adaptive grid.
We extend the adaptive grid technique, known from nonrelativistic
hydrodynamics, to the general relativistic application and identify it
with the concept of shift vectors in a 3+1 decomposition.
The adaptive grid minimizes the number of required
computational zones without compromising the resolution in physically
important regions. Thus, the computational effort is greatly reduced when
the zones are subject to computationally expensive additional
processes, such as Boltzmann radiation transport or a nuclear reaction
network. We present accurate results in the standard tests for
supernova simulations: Sedov's point blast explosion, the
nonrelativistic and relativistic shock tube, the Oppenheimer-Snyder
dust collapse, and homologous collapse.
\end{abstract}



\section{Introduction}

Consider a physical variable \( y \) that evolves in time
\( t \) according to a differential equation \( dy/dt = f(y) \).
The simplest implicit finite differencing
relates the value of \( y^{n} \) at time \( t^{n} \) to its
previous value \( y^{n-1} \) at time \( t^{n-1} \) by
\begin{equation}
\frac{y^{n} - y^{n-1}}{t^{n} - t^{n-1}} = f(y^{n}).
\label{eq_implicit_differencing}
\end{equation}
This equation can be solved for \( y^{n} \) at each time step
\( n=1\ldots\infty \). Depending on the
properties of the function \( f \), a computationally
expensive solution of a nonlinear algebraic equation is required
in most cases.
This is not the case with the explicit
finite differencing where the evaluation of \( f \) is based on the
old values \( y^{n-1} \).
The two discretization variants appear to be similar for small time steps
where \( |f(y^{n}) - f(y^{n-1})| \) is small.
However, the behavior
dramatically differs for larger steps: If a physical equilibrium
with the static solution \( 0 = f(y^{\infty}) \) exists, one easily
checks with equation (\myref{eq_implicit_differencing})
that the equilibrium solution can be approached
in one large time step because of the
bound value of \( |y^{\infty} - y^{n-1}| \).
Explicit finite differencing, on the other hand, does not lead to
a physically meaningful solution with time steps larger than
a characteristic time scale that is determined by the function \( f \).
Implicit methods guarantee
smooth transitions between dynamical phases and equilibrium
phases. On a long time scale, they automatically adjust the
equilibrium to exterior dependencies and provide an
averaging approximation where physical oscillations
around the equilibrium are not of primary interest.
Although implicit schemes certainly are capable of tracking the full
dynamical solution on short time scales, a calculation with time steps
smaller than the characteristic time scale is much more efficiently
done with explicit finite differencing. The reduction in
the number of time steps with the implicit technique
has to compensate the additional
effort invested with the solution of the algebraic equations.
 
Processes in astrophysics involve characteristic
time scales that differ by orders of magnitude. For example
\citet{Henyey_et_al_59} had to introduce
implicit finite differencing in stellar
evolution calculations, where nuclear reactions as well
as hydrodynamics were assumed to be quasi-static on the
time scale of stellar evolution. The system of algebraic equations
is traditionally solved with a Newton-Raphson scheme. The numerical
effort increases steeply with the number of unknowns that have to be
determined in one time step. Thus, implicit approaches are
predestined for systems of moderate size where the evolution time
scale is comparable to the time scale of the underlying physical
processes somewhere in space or time, and exceeds the
latter by far elsewhere or at another time.
Because of this size restriction, implicit hydrodynamics has mainly
been applied in one-dimensional systems, such as in spherically
symmetric hydrodynamics codes for supernova simulations
\citep{Schinder_Bludman_89,Swesty_95,Yamada_97}.
Spherical symmetry additionally allows the use of Lagrangian comoving
coordinates and a full treatment of general relativity
\citep{Misner_Sharp_64}. The evolution of the metric can
be included self-consistently into the solution for the nonlinear
algebraic equations that emerge from the implicit finite differencing.

Fig. (\myref{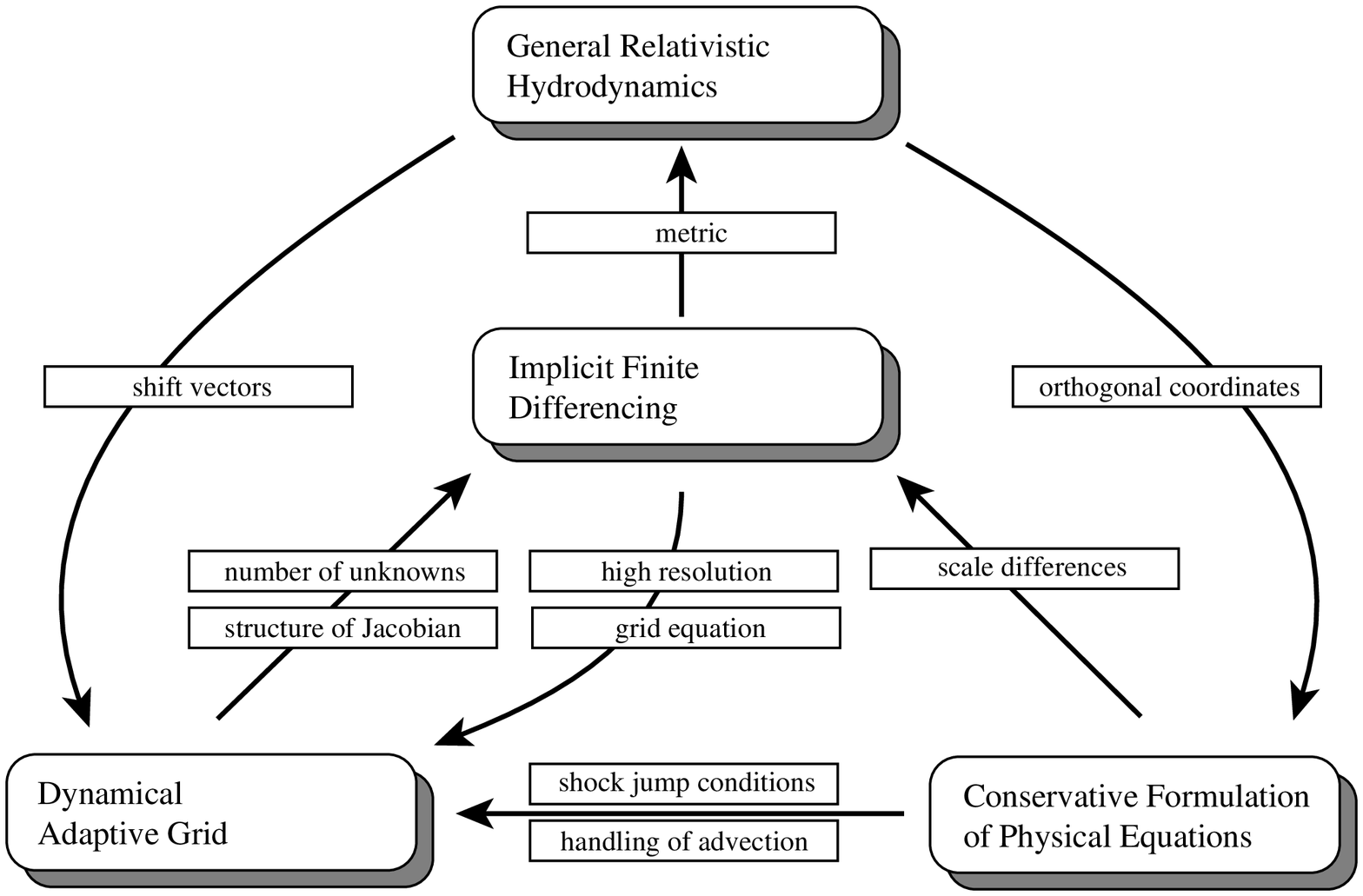}) shows the connection between several
technical features that go along with a hydrodynamics code that
is based on implicit finite differencing.
\citet{Eggleton_71} used an adaptive grid in his quasistatic stellar
evolution code. Together with the hydrostatic equations, a grid
equation is solved that determines where a fixed number of grid
points is located to optimally sample the profile of the star.
A dynamical adaptive grid was suggested by \citep{Harten_Hyman_83}
and the formulation of radiation hydrodynamics on a dynamical adaptive
grid was provided by \citet{Winkler_Norman_Mihalas_84}
and implemented in Newtonian \citep{Winkler_Norman_86}
and special relativistic \citep{Norman_Winkler_86} hydrodynamics.
\citet{Dorfi_Drury_87} found a simple and robust
implicit prescription to guide the dynamics of the adaptive grid
according to the evolution of the physical state. Note the difference
between the {\em dynamical} adaptive grid where a fixed number of
grid points continously moves through the computational domain
and a {\em static} adaptive grid where static grid points are
inserted or removed according to resolution requirements.
The dynamical adaptive grid technique found applications
in astrophysical high-resolution calculations
\citep{Dorfi_Gautschi_89,Mair_90}, and
a well documented implementation of the dynamical adaptive grid technique
for Newtonian hydrodynamics with radiative transfer in different
one-dimensional geometries has been provided by 
\citet{Gehmeyr_Mihalas_94} in the code TITAN.

\begin{figure}
  \plotone{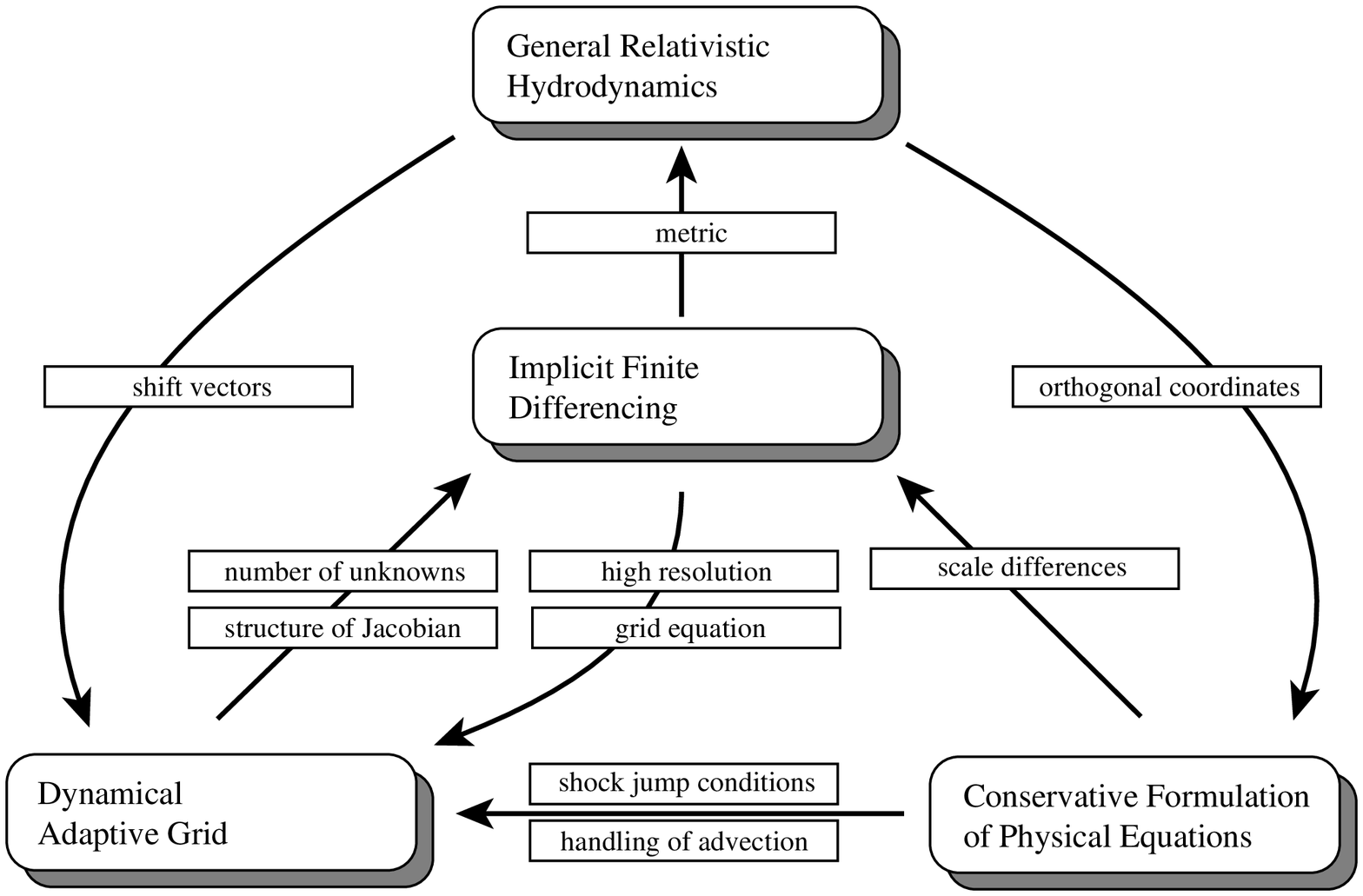}
  \caption{The interdependence between general relativity, implicit
    finite differencing, conservative equations, and the dynamical adaptive
    grid in spherically symmetric hydrodynamics.}
  \mylabel{f1.eps}
\end{figure}
We resume the description of Fig. (\myref{f1.eps}) with the
relationship between a dynamical adaptive grid and implicit hydrodynamics.
The solver of the algebraic equations easily incorporates the solution
of the implicit grid equation. The zones are
dynamically concentrated in the regions of action or where a high
resolution is required. This assures that most of the zones actively
contribute to the interesting part of the evolution and do not clump
together in regions of well-known physical equilibrium.
The dynamical adaptive grid can reduce
the required number of grid points by orders of magnitude
with respect to an equidistant grid, and the implicit scheme greatly benefits
from this size reduction in the solution vector. Since the dynamical
grid changes neither the total number of grid points nor its
ordering, the structure of the Jacobian that has to be inverted in the
Newton-Raphson scheme remains unchanged during the simulation.
Furthermore, the high resolution achieved with the adaptive
grid is only advantageous in combination with implicit hydrodynamics
because the Courant-Friedrichs-Lewy condition
\citep{Courant_Friedrichs_Lewy_28} would impose severe restrictions
on the maximum time step in explicit schemes.

Of course, the adaptive grid also comes with difficulties. The
moving grid relates different spatial points in the physical
state vector across one and the same time step. This requires
a proper handling of advection in the formulation of the physical
equations on the adaptive grid. Moreover, the advection introduces
extended coupling between neighboring zones and therewith increases
the density of the Jacobian in the Newton-Raphson scheme. A conservative
formulation of the physical equation provides a solid basis for
the discretization of the advection. Conserved quantities are
transfered from one zone into a neighboring zone.
A formulation of the dynamics in terms of conservation laws has additional
benefits: (i) Fundamental conservation laws are also valid at
discontinuities and allow an accurate numerical solution - e.g., for
the propagation of shock waves. This is not the case with arbitrary
finite difference approximations. (ii) The integration of a
conserved quantity over adjacent fluid elements does not depend on the
fluxes at the enclosed boundaries.  Thus, discretization errors have
mainly local influence. This advantage is important for the implicit
solution of problems involving scale differences of many orders of
magnitude. (iii) The integration over the whole computational
domain of the single-fluid-element conservation laws leads to the total
conservation of fundamental physical quantities, independent of the
resolution in the conservative finite difference representation.
These contributions of a conservative formulation to the adaptive
grid and implicit solution are also drawn in Fig. (\myref{f1.eps}).

In this paper, we describe our implementation of the dynamical
adaptive grid technique in general relativistic hydrodynamics.
The dynamical adaptive grid is nothing other than a
continuous and time dependent shift in the coordinates with respect
to Lagrangian or Eulerian coordinates. This freedom of coordinate
choice has long been explored in the general relativistic case, where
in addition to the time slices, shift vectors have to be defined in a
3+1 decomposition \citep{Arnowitt_Deser_Misner_62,Smarr_York_78}.
There is no difference between
a dynamical adaptive grid and a specific choice of shift vectors
in the general relativistic description. However, as far as the
numerics is concerned, the formulation with shift vectors only
provides a set of  analytical equations, while the
Newtonian adaptive grid technique can be generalized to
a detailed prescription for a stable numerical implementation
of shift vectors in the general relativistic case.
Thus, instead of directly applying a discretization of appropriate
shift vectors, we first search for a conservative formulation
of hydrodynamics in Lagrangian coordinates and then apply the generalized
adaptive grid technique to implement the shift vectors that lead to
an optimal resolution of the physical state at important locations.

All the features drawn in Fig. (\myref{f1.eps}) come together in an
important astrophysical application: core collapse supernovae. The
compact object at the center of the star requires a general relativistic
treatment \citep{Bruenn_DeNisco_Mezzacappa_01,Liebendoerfer_et_al_01}.
The time scales
of radiation transport, hydrodynamics, and neutrino diffusion differ
by orders of magnitude and call for implicit techniques.
The computational domain covers densities
from \( 10^{5} \) g cm$^{-3}$ in the outer layers of the calculational
domain to several times \( 10^{14} \) g cm$^{-3}$ in
central regions of the proto-neutron star. At these higher-than-nuclear
densities the stiffness of the equation of state is responsible for
very short hydrodynamical time scales. The
speed of sound waves reaches several 10\% of the velocity of light. If
one is not interested in the ringing of the neutron star,
this time scale is prohibitive for the calculation of the long term
physics in the outer range with schemes implementing explicit finite
differencing. The supernova
shock formed after stellar core collapse and bounce requires reasonable
resolution. Moreover, in the case of a supernova explosion, the adaptive grid
allows a smooth propagation of the shock through the outer layers;
and in the case of black hole formation, a robust implementation of
shift vectors might help to shift the arising coordinate singularity
at the Schwarzschild horizon out of the computational domain
\citep{Liebendoerfer_Thielemann_00}.
Even the conservative formulation of the hydrodynamic equations is
physically motivated: A meticulous energy conservation in the finite
difference scheme is required to allow statements on the success of 
a supernova explosion \citep{Mezzacappa_et_al_01}. The balance of
gravitational, internal, and neutrino energy is two orders of magnitude
smaller than the individual constituents.
The total energy is then comparable to the explosion
energy, i.e. about one percent of the energy stored in the neutrino field. The
bookkeeping of this energy has to be based on a careful implementation 
of the mutual energy and momentum transfer between the neutrino radiation 
field and the matter. For
the determination of the explosion energy to an accuracy of at least one
order of magnitude, one may tolerate an energy drift of not
exceeding a fraction of \( 10^{-3} \) of the internal, gravitational or
neutrino radiation energy. This drift extends over the whole
calculation time of more than \( 10000 \) time steps, so that systematic
deviations of energy conservation may not exceed a fraction of \(
10^{-7} \) per time step.
On current computer hardware, it is not yet possible to simulate in
multi-dimensional general relativity the confluence of accurate multigroup
neutrino transport, high resolution hydrodynamics, magnetic fields,
and up-to-date nuclear physics input. The complexity of supernova
explosions is best explored in separate restricted investigations.
Spherically symmetric simulations are able to incorporate the best 
available neutrino transport techniques
\citep{Rampp_Janka_00,Mezzacappa_et_al_01}
in self-consistent simulations
with high resolution and allow the detailed study of high density
weak interactions and nuclear physics.

In the following section, we write down conservative
equations of general relativistic hydrodynamics in spherical
symmetry and discuss the relation
between shift vectors and the adaptive grid. The adaptive grid
technique of \citet{Winkler_Norman_Mihalas_84} is extended to
general relativity and
applied in combination with the grid equation of \citet{Dorfi_Drury_87}.
We then proceed with a description of our approach to
automatically and implicitly integrate a system of stiff ordinary
differential equations in time and outline the
spatially staggered finite differencing of the physical equations.
Numerical diffusion and artificial viscosity are discussed.
In section \myref{section_tests} we run the standard supernova
test simulations with our code, AGILE, and compare the results to
analytical solutions and the Piecewise Parabolic Method (PPM)
\citep{Colella_Woodward_84}.


\section{Conservative Equations on the Dynamical Adaptive Grid}

\subsection{Relativistic hydrodynamics in spherical symmetry}

General relativistic four-dimensional space-time can be described in
terms of consecutive three-dimensional time slices
\citep{Arnowitt_Deser_Misner_62}. The definition of the time
slices and the choice of coordinates within the slices is not given by
the dynamics of the matter; it has to be imposed from the outside as a part
of the description. The time slices comprise the events that are
made simultaneous with respect to coordinate time. They
are defined by a lapse function that describes the proper time lapse
to the infinitesimally close neighbor time slice \( t+\delta t \)
along a path orthogonal to the time slice at time \( t \).
The three-dimensional coordinates within a time slice are then
specified by a shift vector function describing the spatial
shift of points with the same coordinate label between infinitesimally
close neighbor slices. Reference for the shift is again the path orthogonal
to the slice \citep{Smarr_York_78}.

In Newtonian hydrodynamics, time and space are understood as separate
entities. Only one definition of the time slices is possible: orthogonal
to the time axis. However, a free choice of coordinates within the time
slices persists. We may choose Eulerian coordinates along the time axis
or Lagrangian coordinates along fluid element trajectories. If one
understands the grid point labels in a dynamical adaptive grid
technique as coordinates, many mappings between such coordinates
and the location of fluid elements can be realized with appropriate
prescriptions for the motion of the adaptive grid. These choices are
easily interpreted in the general relativistic view: The Eulerian coordinates
correspond to vanishing shift vectors and the Lagrangian coordinates correspond
to shift vectors equal to the three-velocity of fluid elements. In the
adaptive grid case, the shift vectors are identified with the grid velocity.

In the following investigation, the {\em time slices} are chosen
orthogonal to matter trajectories such as to produce comoving coordinates
with {\em vanishing } shift vectors. Nevertheless, the relationship between
adaptive grids in Newtonian space-time and shift vectors in the general
relativistic context provides a guideline for the implementation of shift
vectors in general relativistic simulations. Fixed coordinate intervals
can be mapped to the time slices at will in order to
resolve physically interesting regions with high precision.

We start with the spherically symmetric Einstein equations
in comoving orthogonal coordinates as given by
\citet{Misner_Sharp_64}. They are based on the metric
\begin{equation}
ds^{2}=-\alpha ^{2}dt^{2}+\left( \frac{r'}{\Gamma }\right) ^{2}da^{2}
+r^{2}\left( d\vartheta ^{2}+\sin ^{2}\vartheta d\varphi ^{2}\right) 
, \mylabel{eq_comoving_metric}
\end{equation}
where \( r \) is the areal radius and \( a \) is a label corresponding
to an enclosed rest mass (the prime denotes a derivative with respect
to \( a \): \( r'=\partial r/\partial a \)).  The proper time lapse of
an observer attached to the motion of rest mass is related to the
coordinate time \( dt \) by the lapse function \( \alpha  \).  The
angles \( \vartheta  \) and \( \varphi  \) describe a two-sphere.

The thermodynamic state of a fluid element is given in its
rest frame by the rest mass density, \( \rho \), the temperature,
\( T \), and, in our application, the electron fraction
\( Y_e \). An equation of state renders the composition in nuclear
statistical equilibrium, the specific internal energy, \( e \),
and the isotropic pressure \( p \). In analogy to the definitions
of \citet{Romero_et_al_96} in polar slicing and radial
gauge, we compose conserved quantities for specific volume, total
energy, and radial momentum:
\begin{eqnarray}
\frac{1}{D} & = & \frac{\Gamma }{\rho }\mylabel{eq_specific_volume}\\
\tau & = & \Gamma e+\frac{2}{\Gamma +1}\left( 
\frac{1}{2}u^{2}-\frac{m}{r}\right)
\mylabel{equatino_specific_energy}\\
S & = & u\left( 1+e\right).
\mylabel{eq_specific_momentum}
\end{eqnarray}
The velocity \( u \) 
is equivalent to the \( r\,  \)-component of the fluid four-velocity 
as observed
from a frame at constant areal radius \citep{May_White_67}.
In the special relativistic limit, \( \Gamma =\sqrt{1+u^{2}-2m/r} \)
becomes the Lorentz factor that takes the boost between
the inertial and the comoving observers into account.
The gravitational mass \( m \) is the total energy enclosed 
in the sphere at rest mass \( a \). 
In the nonrelativistic limit, we retrieve with \( \alpha =\Gamma =1 \)
the familiar specific volume \( 1/D = 1/\rho \), the sum of the specific
internal, kinetic, and gravitational energy \( \tau = e + u^2/2 - m/r \), and
the specific radial momentum \( S = u \).

With these definitions, the complete system of general relativistic
hydrodynamics equations can be written in a conservative form
\citep{Liebendoerfer_Mezzacappa_Thielemann_01}:
\begin{eqnarray}
\frac{\partial}{\partial t}\left[ \frac{1}{D}\right]  & = & \frac{\partial 
}{\partial a}\left[ 4\pi r^{2}\alpha u\right] \mylabel{eq_continuity} \\
\frac{\partial \tau }{\partial t} & = & -\frac{\partial }{\partial a}
\left[ 4\pi r^{2}\alpha up \right]
\mylabel{eq_total_energy} \\
\frac{\partial S}{\partial t} & = & -\frac{\partial }{\partial a}
\left[ 4\pi r^{2}\alpha \Gamma p \right] \nonumber \\
 & - & \frac{\alpha }{r}\left[ \left( 1+e+\frac{3p}{\rho }\right) 
\frac{m}{r}+8\pi r^{2} \left( 1+e\right) 
p -\frac{2p}{\rho }\right] \mylabel{eq_momentum} \\
\frac{\partial V}{\partial a} & = & 
\frac{1}{D}\mylabel{eq_volume_gradient} \\
\frac{\partial m}{\partial a} & = & 1+\tau \mylabel{eq_mass_gradient} \\
0 & = & \left( 1+e\right) \frac{\partial \alpha }{\partial 
a}+\frac{1}{\rho }\frac{\partial }{\partial a}\left[ \alpha p\right] 
.\mylabel{eq_lapse_gradient} 
\end{eqnarray}
Equations (\myref{eq_continuity})-(\myref{eq_momentum}) describe the conservation
of volume (analogously to mass conservation in Eulerian coordinates),
total energy, and radial momentum respectively.
The constraints (\myref{eq_volume_gradient})
and (\myref{eq_mass_gradient}) explain themselves in the analogy
to the Newtonian limit, where the first becomes the definition
of the rest mass density and the second the Poisson equation for
the gravitational potential. The enclosed volume in
Eq. (\myref{eq_volume_gradient}) is defined by the areal radius
as in the Newtonian limit, \( V=4\pi r^3/3 \). The condition for the lapse function
in equation (\myref{eq_lapse_gradient}) is derived from the space-space
components of the Einstein field equations.

\subsection{Shift vectors and the dynamical adaptive grid}
\label{section_adaptive_grid}

In Fig. (\myref{f1.eps}) we have now moved from the equations
of general relativistic hydrodynamics to their conservative formulation
in specific
coordinates as shown in the lower right corner. In order to traverse
to the left towards a stable implementation of the adaptive grid,
we rewrite the derivations of 
\citet{Winkler_Norman_Mihalas_84} in a form that applies to the
general relativistic case as well. The goal is a stable and conservative
finite difference representation of equations
(\myref{eq_continuity})-(\myref{eq_momentum}) in terms of \( n \)
grid points, \( (a_{i}(t), i=1\ldots n) \), that continuously
move with respect to the enclosed rest mass label.

We concentrate on observers at rest in their slice. They reside in
local orthogonal reference frames that we assume to be collinear
with global coordinates \( A \). Consequently, coordinates \( A \)
have vanishing shift vectors everywhere. We introduce another system
of global coordinates, \( B \),
having arbitrary shift vectors on the same time slices. Let us
select a fixed grid in coordinates \( B \):
\( \{ \vec{q}_{i\in {\rm N}^3} \in {\rm R}^3,
i_1=1\ldots n_1, i_2=1\ldots n_2, i_3=1\ldots n_3 \} \).
The trace of the world-lines along this grid defines
a moving grid in coordinate system \( A \): \( \{ \vec{a}(t,\vec{q}_{i}) \} \).
The grid world lines along coordinate system \( B \) cut the
continuum into zones \( \Delta q \) on each time slice.
These zones contain a time dependent selection of observers in
coordinate system \( A \): \( \Delta a = \{ \vec{a}(t,\vec{q}), \vec{q}\in \Delta q \} \).
Based on the observations \( y \) of single observers \( \vec{a} \) in
the zone, we define a zone integral of the observations
\begin{equation}
\langle y(t,\vec{a}) \rangle = \int_{\Delta a} y(t,\vec{a}) d\vec{a}.
\end{equation}
In coordinate system \( A \),
the borders of the zones change with time
according to the shift vectors in coordinate system \( B \).
In Newtonian parlance, one can identify the moving zones in coordinate
system \( A \) with the motion of an adaptive grid.
The important point is that the adaptive grid only regroups the observers
into new zones. As stated already by \citet{Winkler_Norman_Mihalas_84},
the adaptive grid never changes the reference frame of the
observations. It solely determines the zonewise support
for the {\em integration} of the observations in the time slice.

In order to formulate time evolution equations, we are interested
in the temporal change of the zone-integrated quantity
\begin{equation}
\langle y(t,\vec{q}) \rangle = \int_{\Delta a} y(t,\vec{a}) d\vec{a}
\end{equation}
in a specific zone \( \Delta q \). We can apply the Reynolds
theorem to relate the time derivative
\( \partial \langle y(t,\vec{q}) \rangle /\partial t \)
in coordinate system \( B \) to the time derivative
\( \langle \partial y(t,\vec{a}) /\partial t \rangle \) in coordinate system \( A \):
\begin{equation}
\frac{\partial \langle y(t,\vec{q}) \rangle}{\partial t} =
\left< \frac{\partial y(t,\vec{a})}{\partial t} \right>
+
\int_{\partial (\Delta a)} y(t,\vec{a}) \left( \frac{\partial \vec{a}(t,q)}{\partial t}
\cdot \vec{n} \right) dS.
\mylabel{reynolds_thorem}
\end{equation}
Note that the Reynolds-theorem is a mathematical relation that has no
physical content. The left hand term is the temporal change in the zone
average of \( y \) in the numerically accessible coordinate system \( B \).
The first term on the right hand side relates this time derivative to
the time derivative in orthogonal coordinates \( A \) that is easily
evaluated based on the physical equations describing the time evolution.
The correction in the second term on the right hand side is due to the
grid motion. It is an integral over the zone surface elements \( dS \)
(with normal \( \vec{n} \)) that corrects for the observations
made by observers that enter or leave the zone during grid motion.

If an equation describing the physical evolution is written in a
conservative form
\begin{equation}
\left< \frac{\partial y(t,\vec{a}) }{\partial t} \right> + 
\int_{\partial (\Delta a)} \left( \vec{f}(t,\vec{a}) \cdot \vec{n} \right) dS =
\int_{\Delta a} s(t,\vec{a}) d\vec{a}
\end{equation}
with a source \( s(t,\vec{a}) \), its formulation on the adaptive grid
simply becomes
\begin{equation}
\frac{\partial \langle y(t,\vec{q}) \rangle }{\partial t} +
\int_{\partial (\Delta a)}
\left( \left[ \vec{f}(t,\vec{a}) - y(t,\vec{a}) \frac{\partial \vec{a}(t,\vec{q})}{\partial t} \right]
\cdot \vec{n} \right) dS = \int_{\Delta a} s(t,\vec{a}) d\vec{a}.
\end{equation}
The motion of zone boundaries naturally leads to a
flux entering or leaving the zone. The rate of coordinate change
at a fixed grid point \( v^{\rm rel} = -\frac{\partial \vec{a}(t,\vec{q})}{\partial t} \)
defines a ``grid velocity'' with respect to coordinates \( A \).
In our spherically symmetric case, the Reynolds theorem reduces to
\begin{equation}
\frac{\partial }{\partial t}\int _{a(t,q_{i})}^{a(t,q_{i+1})}yda=
\int _{a(t,q_{i})}^{a(t,q_{i+1})}\frac{\partial y(t,a)}{\partial t}da+
\left[ y \frac{\partial a(t,q)}{\partial t}\right] _{a(t,q_{i})}^{a(t,q
_{i+1})}
\label{eq_spherical_reynolds}
\end{equation}
and leads for example to the continuity equation (\myref{eq_continuity})
on the adaptive grid
\begin{equation}
\mylabel{adaptive_energy_equation}
\frac{\partial }{\partial t}\int _{a(t,q_{i})}^{a(t,q_{i+1})}\frac{1}{D}da
+\left[ -4\pi r^{2}\alpha u - \frac{1}{D}\frac{\partial a(t,q)}{\partial t}
\right] _{a(t,q_{i})}^{a(t,q_{i+1})} = 0.
\end{equation}
The integral of the conserved quantity (e.g. volume) over
the rest mass of the star only depends on surface terms in this
favorable form because
contributions from zone interfaces exhibit exact cancellation.
Thus, the motion of the adaptive grid affects local resolution
but not global conservation.

\subsection{Dynamics of the adaptive grid}

After having described the formulation of physical equations on an
adaptive grid, we focus on the independent question, how to adapt
the grid to interesting features in the physical evolution. The basic
idea is not only to resolve steep gradients with a locally increased
grid point concentration. To a certain extent, the dynamical adaptive
grid is also capable to capture a self-similar flow and propagate it
through the computational domain mainly by grid motion instead of
rapid changes in the physical state of zones. This can lead to
larger time steps when steep gradients are present.
For example, if a shock front is smeared out over several
grid points, it takes a width \( \Delta q \) in grid point labels \( q
\) (see Fig. (\myref{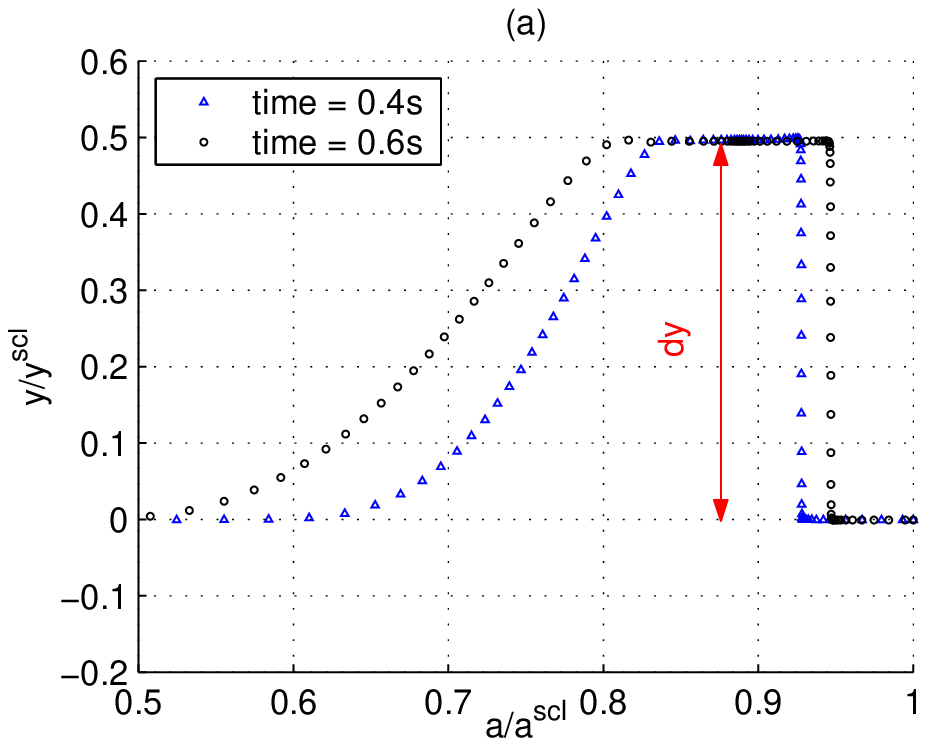})).
\begin{figure}
  \plottwo{f2a.ps}{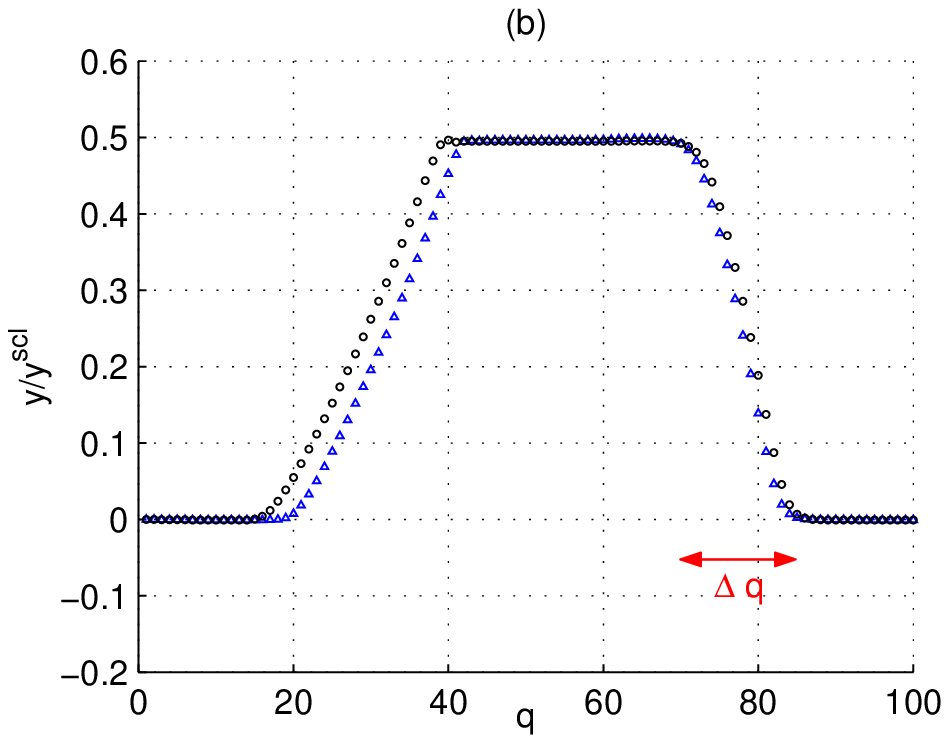}
  \caption{This shock tube simulation (described in section
    \myref{section_PPM}) illustrates the dynamics of the adaptive grid.
    Data at two different times are shown.
    Graph (a) shows the scaled velocity
    profile in coordinates A, i.e. versus enclosed mass. We indicate
    the velocity jump across the shock with \( dy \). The grid is
    set to produce equidistant generalized arclength between grid points.
    This is
    only qualitatively visible in this graph because of the influence
    of the other variables and the spatial smoothing operator that guarantees
    a smooth reduction of the resolution with increasing distance from
    the shock. Graph (b) shows the same profile in coordinates B, i.e.
    versus grid label. The shock is smoothly spread over an interval
    \( \Delta q \) that is used to estimate the time step restriction, equation
    (\myref{shock_time_step}). The profile in the computer representation
    (b) is fairly stationary, while the shock propagation in physical
    space (a) is dominated by the evolution of the mapping \( a(t,q) \).}
  \mylabel{f2a.ps}
\end{figure}
Suppose the shock moves with velocity \( v_{q} \) with respect to
the grid; then there is time \( \Delta t=\Delta q/v_{q} \) to change
the variables from their pre-shock to their post-shock values. We
consider the variable \( y \) with the largest relative jump \( |dy/y|
\) at the shock front. If the relative change of the variable per time
step is limited to \( \varepsilon  \) one gets from \( |y|+|dy|\leq
|y|(1+\varepsilon )^{n} \) about \( n=\ln (1+|dy/y|)/\ln (1+\varepsilon
) \) time steps for this change. Therefore, the time step is limited by
\begin{equation}
\mylabel{shock_time_step}
\Delta t<\frac{\Delta q\ln (1+\varepsilon )}{v_{q}\ln (1+|dy/y|)}.
\end{equation}
In contrast to Lagrangian or Eulerian schemes, this time limit is not
restrictive with the adaptive grid. Firstly, the high resolution
at the shock front gives a large \( \Delta q \) and secondly, the fact
that the grid points tend to move with the shock structure makes the
velocity difference between the shock front and the grid, \( v_{q} \),
very small.

We have implemented the grid equation of
\citet{Dorfi_Drury_87}. We only differ in one detail: The scheme has been
transferred from Eulerian coordinates in flat space to comoving
coordinates in Misner-Sharp time slices. Thus, we retrieve a
Lagrangian scheme if the adaptive grid is switched off.
The grid equation of Dorfi and Drury is based on a resolution
function,
\begin{equation}
R(t,a) = \left( 1 + \sum \left( \frac{a^{{\rm scl}}}{y^{{\rm scl}}}
\frac{\partial y(t,a)}{\partial a} \right) ^2 \right) ^{\frac{1}{2}},
\end{equation}
where the sum includes a selection of relevant variables
\( y \) as e.g. velocity, density, etc.. We set the global scale
in this equation to
\( y^{\rm scl} = \sum_{i} \left|y_{i+1}-y_{i} \right| \).
In terms of a grid point concentration
\( n = a^{{\rm scl}}(\partial a(t,q)/\partial q)^{-1} \)
the grid equation implements the requirement
\begin{equation}
\frac{\partial}{\partial q}\left( \frac{n}{R} \right) = 0.
\mylabel{eq_adaptive_grid}
\end{equation}
If we imagine the variables \( y/y^{{\rm scl}} \) plotted in
a graph versus \( a/a^{{\rm scl}} \), this condition enforces
a constant value of
\begin{equation}
\frac{n}{R} = \left(
\left(\frac{\partial a(t,q)}{a^{{\rm scl}}\partial q} \right) ^2
+ \sum \left(\frac{\partial y(t,q)}{y^{{\rm scl}}\partial q} \right) ^2
\right) ^{-\frac{1}{2}},
\end{equation}
equivalent to a constant generalized arc length per grid point
interval \( \Delta q \). The projection of grid points
to the \( a/a^{{\rm scl}} \)-axis results in an increased grid point
concentration at locations with steep gradients in the variables
\( y/y^{{\rm scl}} \). This basic approach is refined by the
sophisticated application of operators for spatial smoothing
and temporal retardation of the grid adaption
as described by \citet{Dorfi_Drury_87} (see Fig. (\myref{f2a.ps})).


\section{Numerical Method}

The nonlinear partial differential equations
(\myref{eq_continuity})-(\myref{eq_momentum}) and the
constraints (\myref{eq_volume_gradient})-(\myref{eq_lapse_gradient}),
(\myref{eq_adaptive_grid}) are finite differenced
in the spatial direction to build a large system of
coupled ordinary differential equations with algebraic
constraints. Each independent variable
in each zone acts as an individual unkown function
of time in the solution vector. In a second step, this
large system of potentially stiff algebro-differential
equations is integrated in time with our solver AGILE.
Experimentation was necessary to find the final finite
difference representation of the hydrodynamical equations.
We have therefore separated the solver for the
ordinary differential equations and the adaptive grid
equation from the modules that describe and discretize
the actual physical equations. AGILE has the
capapility to calculate the Jacobian of the physical
equations automatically for the implicit time evolution.
This feature substitutes the cumbersome and errorprone
coding of the Jacobian and leaves room for the motivation
to change and improve the physical equations.

In the following subsection we start with the description
of the time evolution in AGILE. Then, we discuss the
numerical diffusion induced by the advection that accompanies
the adaptive grid. We introduce artificial viscosity and
review the role it plays in combination with the adaptive
grid technique. Finally, we outline our complete spatial
discretization of the Einstein equations in comoving coordinates.

\subsection{Time evolution in AGILE}

We start with a vector, \( y \), that depends on the
parameter time, \( t \). The temporal evolution of the vector \( y
\) shall be described by an implicit system of equations \begin{equation} F\left(
t,y,\frac{\partial y}{\partial t}\right) =0.\end{equation}
The set of equations \( F \) is assumed to contain the same number of
equations as the vector \( y \) contains components to be solved for.
In the discretization of these equations we use upper
indices to denote the time steps: \( y^{n} \) is a vector of the variables
at time \( t^{n} \) and \( y^{n+ 1} \) is the vector after the time
step \( dt=t^{n+1}-t^{n} \). The discretized equations then have
the form \( F\left( y^{n},y^{n+1};dt\right) =0 \), where the time
derivative of the state vector \( y \) is calculated by a finite
difference representation included in the equations \( F \).

A time step \( dt \) is chosen such that the relative change in
the components of the state vector is not expected to exceed a given value
in a single time step. The solution
vector \( y^{n+1} \) at time \( t^{n+1}=t^{n}+dt \) is then found by a
Newton-Raphson scheme: The equations are Taylor expanded around a
guessed solution vector \( \tilde{y}^{n+1 } \)
\begin{equation}
\mylabel{eq_implicit_taylor_expansion}
F(y^{n},\tilde{y}^{n+1}+\Delta y;dt)=F(y^{n},\tilde{y}^{n+1};dt)+\frac{\partial 
F(y^{n},\tilde{y}^{n+1};dt)}{\partial y^{n+1}}\Delta y+O((\Delta y)^{2}).
\end{equation}
and solved for the corrections
\begin{equation}
\mylabel{eq_implicit_step}
\Delta y=-\left( \frac{\partial F(y^{n},\tilde{y}^{n+1};dt)}
{\partial y^{n+1}}\right) ^{-1}F(y^{n},\tilde{y}^{n+1};dt).
\end{equation}
This procedure is iterated until a norm of the correction vector
\( |\Delta y|/y^{n,{\rm scl}} \) becomes sufficiently small. We
use the scaling \( y^{n,{\rm scl}}=|y^{n}|+y_{min}^{{\rm scl}}
\). The first term usually dominates and the second term limits
the sensitivity to numerical noise in a very small component.

The chosen notation does not show at which time between \( t^{n} \)
and \( t^{n+1} \) the equations actually are solved. In the fully
implicit approach, the old values \( y^{n} \) are only used in the time
derivatives and the equations are solved at \( t^{n+1} \). In the
literature one also finds semi-implicit approaches
\citep{Winkler_Norman_86,Mair_90,Swesty_95}
that solve the equations at a time \( t^{n+\lambda } \)
depending on some averaged variables \( y^{n+\lambda }(y^{n},y^{n+1})
\) with \( 0<\lambda <1 \). For \( \lambda =0 \) one obviously gets an
explicit scheme. Optimally, with \( \lambda =1/2 \), a semi-implicit
scheme is achieved whose time discretization error is second order in
the time step. Unfortunately this choice is often not stable enough.
A compromise are schemes with \( \lambda \simeq
0.7 \) which is reported to be about the minimum \( \lambda  \) with
satisfying stability \citep{Winkler_Norman_86}.
For the time being, we use the fully implicit choice \( \lambda =1 \)
because this setting is sufficient for the supernova application.
However, we experimented with a semi-implicit multi-step
extrapolation method that generalized the approach of
\citet{Bader_Deuflhard_83} to algebro-differential
equations \citep{Liebendoerfer_Rosswog_98,Liebendoerfer_00}.
The great benefit from the higher order extrapolation in combination
with the smooth
computer representation of the physical state on the adaptive grid was,
however, severely disturbed by the discontinuous switches in the
upwind advection scheme. We therefore
focus on the simple and stable fully implicit integration and
revive our higher order integration scheme in AGILE
({\bf A}daptive
{\bf G}rid with {\bf I}mlicit {\bf L}eap {\bf E}xtrapolation)
at a later occasion.

We implemented the following automatic
procedure that calculates the Jacobian numerically based
on the modules containing the physical equations.
For the first iteration, we assume that we have a guess
\( \widetilde{y}^{n+1} \)
(e.g. \( \widetilde{y}^{n+1} = y^{n} \)) at hand for the
state vector at the new time and that we know the sparsity
pattern of the Jacobian well enough to specify a region that
contains all nonzero coefficients (e.g. a large band).
At the old time \( t^{n} \) we then
attribute a scale \( y^{n,{\rm scl}} \) to the state vector \( y^{n} \)
and evaluate the residuum vector \(
\widetilde{R}=F(y^{n},\widetilde{y}^{n+1};dt) \) based on the guess.
For the calculation of the numerical derivatives, we define a
small number \( \varepsilon  \) based on machine precision.
Furthermore, the components of the state vector, denoted with
label \( i \), are sorted into \( J \) distinct groups \( g_{j} \)
according to the rule that none of the equations \( F \) may depend
on more than one state vector component out of the same group.
Based on the sparsity
pattern, the groups are formed such as to minimize the required
number of distinct groups.  We select a group \(
g_{j} \) and create a copy \( \widetilde{y}^{var,n+1}\simeq
\widetilde{y}^{n+1} \) of the guess with a small variation
\begin{eqnarray}
\widetilde{y}[i]^{var,n+1} & = & \widetilde{y}[i]^{n+1}+\varepsilon y[i]^{n,{\rm scl}}
\nonumber \\
\Delta [i] & = & \frac{\widetilde{y}[i]^{var,n+1}-\widetilde{y}[i]^{n+1}}{y[i]^{
n,{\rm scl}}}
\end{eqnarray}
in all components \( i\in g_{j} \). The residuum vector \(
\widetilde{R}^{var}=F (y^{n},\widetilde{y}^{var,n+1};dt) \) is then
calculated with the varied guess. From the two residua we
can extract the components of the Jacobian
\begin{equation}
A[k,i]=y[i]^{n,{\rm scl}}\frac{\partial
F[k](y^{n},\tilde{y}^{n+1};dt)}{\partial
y[i]^{n+1}}=\frac{\widetilde{R}[k]^{var}-\widetilde{R}[k]}{\Delta
[i]}\end{equation} for all \( i\in g_{j} \). The Jacobian is complete when
this procedure has been performed for all groups \( J \). Finally, we scale
the rows of \( A \) and the right hand side residuum vector by the maximum
component in the row and solve the linear system
\begin{equation}
\sum_{i}A[k,i]\left( \Delta y[i] / y[i]^{n,{\rm scl}} \right) =
-\widetilde{R}[k].
\end{equation}
to get the corrections to the guess.
Before the next iteration, all zeros in the Jacobian are detected
and the sparsity structure is refined. This allows
further reduction of the number of required groups for all
following time steps. The evaluation of \(
\widetilde{R}^{var}=F(y^{n},\widetilde{y}^{var, n+1};dt) \) for a group
\( j \) is completely independent from the corresponding calculation
for a different group \( j'\neq j \). In order to compose a complete
Jacobian, the system of equations has to be calculated once with the
actual guess, and \( J \) times in parallel with varied guesses.

This parallelism, in principle, allows the build of the Jacobian
on parallel machines without affecting the modules with the physical
equations. The whole system of equations is still evaluated at once.
The physical equations for the different components in the state
vector can therefore make use of common subexpressions or be vectorized.
If for each group a separate process is available,
the described parallelism maximally reduces the wall-clock time per
implicit time step to the single process time of the evaluation for
one residuum vector as in an explicit scheme. However, this comparison
does not include the time spent with the inversion of the Jacobian.
A realistic estimate of the gain in wall clock time has to await
a physical need to complete the implementation of this feature. Note also
that the described separation into groups is not efficient with an
arbitrary sparse structure. It however is efficient with the block-diagonal
Jacobian occurring in spherically symmetric hydrodynamics.

\subsection{Diffusion and artificial viscosity}
\label{section_diffusion}

Before describing a detailed finite difference representation of the
Einstein equations (\myref{eq_continuity})-(\myref{eq_lapse_gradient})
in the next section,
we stress here a few discretization principles that were useful
in supernova simulations, and might carry over to other applications
as well.
The modeling of core collapse supernovae presents a special challenge for the
finite differencing of the hydrodynamics equations. On the one hand,
{\em global} energy conservation
is important because of the comparable size of
the gravitational, internal, and neutrino radiation energy.
The balance between those defines the two orders of magnitude
smaller total energy. The latter is comparable to the expected
explosion
energy and has to be accurately evolved in order to allow conclusions
on the explosion energy in the model. On the other hand, the accurate
evolution of the {\em internal} energy and temperature is essential for
the strongly temperature-sensitive
weak interaction cross sections that figure as a prerequisit to
neutrino transport, which may determine the outcome of a supernova
simulation. Thus, one
would like to solve an equation for the evolution of global energy
to obtain exact energy conservation, and an equation for the evolution of
internal energy to obtain accurate temperatures, and a momentum equation
to determine accurate fluid velocities that relate to the kinetic energy.
However, this set of equations is overdetermined since the total
energy (without gravitation) is nothing more than the sum of internal
and kinetic energy. The following measures were necessary to obtain
good results in all three quantities with the solution of only two
independent equations:

(i) In the Newtonian limit, the time derivative of the specific
total energy, $\tau=e+1/2u^2$, relates its evolution to the evolution
of specific internal energy, $e$, and the evolution of specific
linear momentum (=velocity), $u$. Consequently, the finite
difference representation of the internal energy equation and
the momentum equation determine a consistent finite difference
representation for the total energy equation. The challenge is
to find a discretization of these constituents such that energy
conservation becomes manifest in the finite difference
equation for total energy.
By making the three evolution equations consistent in their
finite difference representation, it is less important, which two equations
are independently evolved and which third quantity is determined
by them.
 
(ii) However, we found an exact match only for the terms up to order
$(v/c)$. Moreover, the match does not include the effect of the
advection terms, and truncation errors may
swamp the solution of the dependent quantity when it happens to
be generated out of the cancellation of two much larger independently
evolved quantities (e.g. the evolution of total energy and large kinetic
energy lead to a poor extraction of internal energy). In order to
keep these inaccuracies small, we pick the evolution of total energy 
as our first independent equation and complement it with the evolution
of the ratio of kinetic and internal energy as second independent
equation. The time derivative of this ratio,
\begin{equation}
\frac{e^2}{u}\frac{\partial}{\partial t} \left( \frac{u^2}{2e} \right)
= e\frac{\partial u}{\partial t} - \frac{1}{2}u\frac{\partial e}{\partial t},
\end{equation}
is a linear combination of the internal energy equation and the momentum
equation that favors the evolution equation of the smaller quantity.
We consider global energy conservation and accurate internal
energies to be more important in a supernova simulation than
the lost strict conservation of radial momentum.
 
(iii) The finite difference representation of the dynamical equations
introduces numerical diffusion. Part of this diffusion is desired to
stabilize the solution. E.g. the upwind differencing in the momentum
advection scheme acts as artificial diffusion, dissipates kinetic energy,
and consequently leads to a mismatch in the relation between total,
internal, and kinetic energy evolution. However, the artificial diffusion
introduced
by upwind differencing can be evaluated and related to a source term
for the internal energy equation. The dissipated kinetic energy is then
correctly transformed to internal energy that restores overall consistency.
We will include this diffusive advection term into the concept of artificial
viscosity that is needed in our scheme to limit the resolution
of a shock front. The artificial viscosity tensor is based on physical
viscosity and consistently included in the general relativistic set
of hydrodynamics equations. It provides the desired mechanism for
dissipation of kinetic energy into internal energy. A more detailed
discussion of the artificial diffusion and viscosity is the aim of this
section.

Along the lines of equation (\ref{eq_spherical_reynolds}), a generic time evolution
equation for a conserved quantity \( y \) would be finite differenced as
\begin{equation}
\frac{y_{i'}da_{i'}-\overline{y}_{i'}\overline{da}_{i'}}{dt}+y^{{\rm adv}}_{i+1}-y^{{\rm adv}}_{i}-y^{{\rm ext}}_{i'}=0.\end{equation}
Here we denote the spatial dependence of the quantities by zone edge indices
$i$ and adress a zone center value by the index $i'=i+1/2$.
Quantities with an overbar belong to the past time \( t \) while
all other quantities belong to the present time \( t+dt \). The temporal change
of the quantity \( y_{i'}da_{i'} \) in a zone with extent \( da_{i'} \)
is given by the advection at its
boundaries and a not further specified external source \( y^{{\rm ext}}_{i'} \).
Let us set the advection terms by first order upwind differencing depending
on the direction of the relative velocity \( u_{i}^{\rm rel} \) defined in
equation (\myref{eq_grid_velocity}):
\begin{equation}
y^{{\rm adv}}_{i}=\left\{ \begin{array}{ll}
y_{i'-1}u^{{\rm rel}}_{i} & {\rm if}\quad u^{{\rm rel}}_{i}\geq 0,\\
y_{i'}u^{{\rm rel}}_{i} & {\rm otherwise}.\end{array}\right.
\end{equation}
We would obtain an identical advection scheme by writing
\begin{equation}
y^{{\rm adv}}_{i}=\frac{1}{2}u^{{\rm rel}}_{i}\left( y_{i'}+y_{i'-1}\right) -\frac{1}{2}\left| u^{{\rm rel}}_{i}\right| \left( y_{i'}-y_{i'-1}\right). \end{equation}
Hence, we may interprete it as the superposition of the second order accurate interpolated
advection \( u^{{\rm rel}}_{i}\left( y_{i'}+y_{i'-1}\right) /2 \) plus a correction
term \( -\left| u^{{\rm rel}}_{i}\right| \left( y_{i'}-y_{i'-1}\right) /2 \).
It is well-known that the second order term alone does not lead to a monotonic
and stable advection scheme. The additional stabilizing correction term is a
diffusive flux that corresponds to a diffusivity
\begin{equation}
D^{{\rm adv}}=\frac{1}{2}\left| u^{{\rm rel}}\right| da.
\label{eq_advective_diffusivity}
\end{equation}
We therewith note that the advection (related to the dynamical adaptive
grid) introduces a diffusivity that is proportional to the velocity of the grid
with respect to matter, and proportional to the grid spacing. In the case of
the momentum equation, this diffusivity reduces the kinetic energy and consequently
leads to internal energy generation if conservation of total energy is supposed
to hold. In the adiabatic Newtonian limit, only the internal energy equation 
\begin{equation}
\frac{\partial e}{\partial t}+\left( p+\frac{1}{2}\left| u^{{\rm rel}}\right| da\right) \frac{\partial }{\partial t}\left( \frac{1}{\rho }\right) = 0 \end{equation}
complements the momentum equation consistently. In the finite
difference expression of the general relativistic momentum equation we
take this effect into account by not directly coding upwind differences.
Almost equivalently, we use the
second order advection term in equation (\myref{eq_fd_momentum_advection}) and
add the stabilizing diffusion term (\myref{eq_advective_diffusivity}) to
the artificial viscosity tensor (\myref{eq_fd_artificial_viscosity}).
Its role as source for internal energy
is then automatically accounted for by the entry of viscous pressure in
the energy equations. The implementation conceptually mutates
from a diffusive implementation of ideal fluid dynamics to an ``ideal''
implementation of diffusive fluid dynamics.

Artificial viscosity is needed to handle discontinuities in the
solution that may occur in ideal fluid dynamics,
e.g. in shock waves. In a simple finite difference scheme on an equidistant
grid, numerical oscillations appear as soon as the steepness of the shock front
exceeds the resolution of the grid \citep{Noh_78}. These oscillations
can be suppressed
by the introduction of artificial viscosity that limits the steepness of the
shock wave to the resolution that is provided by the grid
\citep{VonNeumann_Richtmyer_50}. The adaptive grid
puts the same concept into a slightly different light: The grid adapts according
to the steepening gradient and is always able to resolve it properly without
initially leading to unphysical oscillations. However, with increasing resolution,
the mass in the zones at the shock front approaches zero and the system of equations
becomes ill-conditioned. Poor convergence of the implicit step is the immediate
response. Artificial viscosity is needed in our context to limit the gradient
at the shock front and prevent the grid resolution from growing indefinitely.

In spherical symmetry, however, the formulation of artificial viscosity
has to be chosen with care to circumvent systematic artificial
heating in phases of homologous compression. We extend the approach of
\citet{Tscharnuter_Winkler_79} to general
relativistic hydrodynamics \citep{Liebendoerfer_Mezzacappa_Thielemann_01}
and define the viscous pressure
based on a parameter \( \Delta l \) with the dimension of a shock width,
\begin{eqnarray}
Q & = & \left\{ \begin{array}{ll}
\Delta l^{2}\rho \mbox {div}(u)\left[ \frac{\partial u}{\partial r}-\frac{1}{3}\mbox {div}(u)\right]  & \mbox {if\quad div}(u)<0,\\
0 & {\rm otherwise} \end{array}\right. 
\label{eq_artificial_viscosity} \\
\mbox {div}(u) & = & \frac{\partial }{\partial V}\left( 4\pi r^{2}u\right) .
\nonumber
\end{eqnarray}
 The corresponding viscous heating in the energy equation becomes: 
\begin{eqnarray}
\frac{\partial e}{\alpha \partial t} & = & \left( \frac{u}{r}-\frac{\partial u}{\partial r}\right) \frac{Q}{\rho } \nonumber \\
 & = & -\frac{3}{2}\left[ \frac{\partial u}{\partial r}-\frac{1}{3}\mbox {div}(u)\right] \frac{Q}{\rho }.
\end{eqnarray}
The first expression shows that the viscous heating vanishes in the case of
homologous compression \( (u/r=\partial u/\partial r) \).
The second expression together with equation
(\myref{eq_artificial_viscosity})
shows that viscous heating always has positive sign. In our system of equations,
the viscosity affects the equation for the total energy evolution (\myref{eq_total_energy}),
the momentum evolution (\myref{eq_momentum}), and the constraint for the lapse
function (\myref{eq_lapse_gradient}). These equations become
\citep{Liebendoerfer_Mezzacappa_Thielemann_01}
\begin{eqnarray}
\frac{\partial \tau }{\partial t} & = & -\frac{\partial }{\partial a}\left[ 4\pi r^{2}\alpha u\left( p+Q\right) \right] \nonumber \\
\frac{\partial S}{\partial t} & = & -\frac{\partial }{\partial a}\left[ 4\pi r^{2}\alpha \Gamma \left( p+Q\right) \right] \nonumber \\
 & - & \frac{\alpha }{r}\left[ \left( 1+e+\frac{3\left( p-Q\right) }{\rho }\right) \frac{m}{r}+8\pi r^{2}\left( 1+e\right) \left( p+Q\right) -\frac{2p}{\rho }+\frac{Q}{\rho }\right] \nonumber \\
0 & = & -\left( 1+e\right) \frac{\partial \alpha }{\partial a}-\frac{1}{\rho }\frac{\partial }{\partial a}\left[ \alpha p\right] -\frac{1}{V\rho }\frac{\partial }{\partial a}\left[ V\alpha Q\right] .
\end{eqnarray}

\subsection{Discretization of the Einstein equations}

In all the other sections, the velocity of light, \( c \),
and the gravitational constant, \( G \), were absorbed in
the choice of the units. For clarity, we include these constants
explicitly in this section, where the details of the finite
differencing are given.
As before, we denote by \( i \) the zone edges and
by \( i'=i+1/2 \) the zone centers.
Quantities that are evaluated at the old time \( t^{n} \) are marked with an
overbar, all other quantities belong to the new time \( t^{n+1}=t^{n}+dt \). The
state vector \( y=\left\{ a,r,u,m,\rho ,T,Y_{e},\alpha \right\}  \) contains
the independent variables: enclosed baryon mass, radius, velocity, enclosed
gravitational mass, rest mass density, temperature, electron fraction, and lapse
function, respectively. From the equation of state we get the pressure \( p \)
and specific internal energy \( e \). First, we specify the enclosed volume
and its time derivative
\begin{equation}
V_{i}=\frac{4\pi r_{i}^{3}}{3},\qquad w_{i}=4\pi r^{2}_{i}u_{i},\end{equation}
and define zone differences for the radius and enclosed rest mass
\begin{equation}
dr_{i'}=r_{i+1}-r_{i},\qquad da_{i'}=a_{i+1}-a_{i}.\end{equation}
Further, we define
\( \Gamma _{i}=\sqrt{1 + (u_{i}/c)^{2} - 2Gm_{i}/(c^2 r_{i})} \) on zone
edges and prepare the conserved ``density,'' \( D_{i'} \), and the ``radial
momentum,'' \( S_{i} \). We split the conserved energy into ``internal'' energy,
\( \tau 1_{i'} \), at the zone centers and ``kinetic'' and ``gravitational'' energy,
\( \tau 2_{i} \), \( \tau 3_{i} \), at the zone edges
\begin{eqnarray}
D_{i'} & = & \frac{\rho _{i'}}{\Gamma _{i'}} \nonumber \\
S_{i} & = & \left( 1+\frac{e_{i}}{c^2} \right) u_{i} \nonumber \\
\tau 1_{i'} & = & \Gamma _{i'}e_{i'} \nonumber \\
\tau 2_{i} & = & \frac{u^{2}_{i}}{\Gamma _{i}+1} \nonumber \\
\tau 3_{i} & = & \frac{2}{\Gamma _{i}+1}\frac{Gm_{i}}{r_{i}}.
\end{eqnarray}
We interpolate \( \Gamma  \) to the zone centers and \( e \) to zone edges
by arithmetic means. Based on the derivation in section \myref{section_adaptive_grid} and
equation (\myref{eq_spherical_reynolds}),
the adaptive grid causes advection of conserved quantities through the zone
boundaries. We denote the relative velocity between zone edges and matter
with
\begin{equation}
u_{i}^{{\rm rel}}=-\frac{a_{i}-\overline{a}_{i}}{dt}
\label{eq_grid_velocity}
\end{equation}
and implement first order upwind differencing for the advection terms. On the
zone edges this leads to
\begin{eqnarray}
V_{i}^{{\rm adv}} & = & \frac{1}{D_{i'-1}}u^{{\rm rel}}_{i} \nonumber \\
\tau 1_{i}^{{\rm adv}} & = & \tau 1_{i'-1}u^{{\rm rel}}_{i} \nonumber \\
e_{i}^{{\rm adv}} & = & e_{i'-1}u^{{\rm rel}}_{i} \nonumber \\
Y_{e,i}^{{\rm adv}} & = & Y_{e,i'-1}u^{{\rm rel}}_{i}
\end{eqnarray}
if \( u^{{\rm rel}}_{i}\geq 0 \) and to
\begin{eqnarray}
V_{i}^{{\rm adv}} & = & \frac{1}{D_{i'}}u^{{\rm rel}}_{i} \nonumber \\
\tau 1_{i}^{{\rm adv}} & = & \tau 1_{i'}u^{{\rm rel}}_{i} \nonumber \\
e_{i}^{{\rm adv}} & = & e_{i'}u^{{\rm rel}}_{i} \nonumber \\
Y_{e,i}^{{\rm adv}} & = & Y_{e,i'}u^{{\rm rel}}_{i}
\end{eqnarray}
otherwise. At the zone centers we set \( u^{{\rm rel}}_{i'}=(u^{{\rm rel}}_{i+1}+u^{{\rm rel}}_{i})/2 \)
and advect the quantities
\begin{equation}
\tau 2^{{\rm adv}}_{i'} = \tau 2_{i}u^{{\rm rel}}_{i'}
\end{equation}
if \( u^{{\rm rel}}_{i'}\geq 0 \) and
\begin{equation}
\tau 2^{{\rm adv}}_{i'} = \tau 2_{i+1}u^{{\rm rel}}_{i'}
\end{equation}
otherwise. For the advection of gravitational energy and momentum, we
set
\begin{eqnarray}
\tau 3^{{\rm adv}}_{i'}=\frac{1}{2}\left( \tau 3_{i+1}+\tau 3_{i}\right) u^{{\rm rel}}_{i'} \nonumber \\
S^{{\rm adv}}_{i'} = \frac{1}{2}\left( S_{i+1}+S_{i}\right) u^{{\rm rel}}_{i'}.
\label{eq_fd_momentum_advection}
\end{eqnarray}
Since the gravitational energy is a rather smooth function, that depends
only weakly on the immediate solution of the energy and momentum equation, we
can implement the more accurate, direction independent, second order advection
term. The radial momentum advection, however, is not smooth. But we prefer
to split the upwind differencing into second order advection and a diffusive
term that is included in the artificial viscosity. The latter is based on
the velocity divergence
\begin{eqnarray}
dV_{i'} & = & \frac{4\pi }{3}dr_{i'}\left( r^{2}_{i+1}+r^{2}_{i}+r_{i+1}r_{i}\right) , \nonumber \\
{\rm div}u_{i'} & = & \min \left( 0,\frac{w_{i+1}-w_{i}}{dV_{i'}}\right)
\end{eqnarray}
and is given by equation (\ref{eq_artificial_viscosity})
and (\ref{eq_advective_diffusivity})
\begin{eqnarray}
Q_{i'} & = & \Delta l^{2}\rho _{i'}{\rm div}u_{i'}\left( \frac{u_{i+1}-u_{i}}{dr_{i'}}-\frac{1}{3}{\rm div}u_{i'}\right) 
\label{eq_fd_artificial_viscosity} \\
& - & \frac{1}{2}\left| u^{{\rm rel}}_{i'} \right|
\left( S_{i+1}-S_{i} \right) \frac{dr_{i}}{\alpha_{i'}\Gamma_{i'}dV_{i'}}
\nonumber \\
e^{Q}_{i'} & = & -\frac{3}{2}\left( \frac{u_{i+1}-u_{i}}{dr_{i'}}-\frac{1}{3}{\rm div}u_{i'}\right)
\frac{Q_{i'}}{\rho_{i'}}da_{i'}.
\nonumber
\end{eqnarray}

After these preparations, we begin with the finite differencing of the
constraints from equations (\myref{eq_volume_gradient})-(\myref{eq_lapse_gradient}).
Straightforward is the discretization of the definition
of the density and the Poisson equation
\begin{eqnarray}
da_{i'}-D_{i'}dV_{i'} & = & 0
\label{eq_fd_volume_gradient} \\
m_{i+1}-m_{i}+\Gamma _{i'}\left( 1+\frac{e_{i'}}{c^2}\right) da_{i'} & = & 0.
\label{eq_fd_mass_gradient}
\end{eqnarray}
Less straightforward is the implementation of the constraint for the lapse function
that is in some details arbitrarily chosen to become
\begin{eqnarray}
\alpha _{i'}p_{i'}-\alpha _{i'-1}p_{i'-1} & + & \frac{1}{V_{i}}\left( V_{i'}\alpha _{i'}Q_{i'}-V_{i'-1}\alpha _{i'-1}Q_{i'-1}\right) \nonumber \\
- \frac{1}{4\pi r_{i}^{2}} \alpha_{i'-1} S_{i}^{\rm ext}
& + & \rho _{i}c^2\left( 1+\frac{e_{i}}{c^2}\right) \left( \alpha _{i'}-\alpha _{i'-1}\right) =0,
\label{eq_fd_lapse_gradient}
\end{eqnarray}
where the zone center values of \( V \) and the zone edge values of the density
\( \rho  \) and internal energy \( e \) are found by arithmetic means.

We can then start with the construction of the time evolution equations. The
continuity equation lives on zone centers and is discretized as 
\begin{equation}
\mylabel{eq_fd_continuity}
\frac{dV_{i'}-\overline{dV}_{i'}}{dt}+V_{i+1}^{{\rm adv}}-V_{i}^{{\rm adv}}-\left( \alpha _{i+1}w_{i+1}-\alpha _{i}w_{i}\right) =0.
\end{equation}
 The equation for internal energy evolution also lives on zone centers and reads
\begin{eqnarray}
F^{e}_{i'} & = & \frac{e_{i'}da_{i'}-\overline{e}_{i'}\overline{da}_{i'}}{dt}+e_{i+1}^{{\rm adv}}-e_{i}^{{\rm adv}}\nonumber \\
 & + & \frac{\alpha _{i'}p_{i'}}{\Gamma _{i'}}\left( w_{i+1}-w_{i}\right) -\alpha_{i'}e^{Q}_{i'}-\alpha_{i'}e_{i'}^{{\rm ext}}=0. 
\mylabel{eq_fd_internal_energy}
\end{eqnarray}
We introduced an additional energy source \( e_{i'}^{{\rm ext}} \) that describes
energy exchange with external processes as e.g. with a nuclear reaction network
or a radiation field. Analogously, we also include an external stress \( S^{{\rm ext}}_{i} \)
and an external compositional change \( Y^{{\rm ext}}_{e,i'} \). The momentum
equation lives on zone edges and is discretized as
\begin{eqnarray}
F^{S}_{i} & = & \frac{S_{i}da_{i}-\overline{S}_{i}\overline{da}_{i}}{dt}
+S_{i'}^{{\rm adv}}-S_{i'-1}^{{\rm adv}} \nonumber \\
 & + & \frac{3}{r_{i}}\left[ V_{i}\left( \Gamma _{i'}\alpha _{i'}p_{i'}
-\Gamma _{i'-1}\alpha _{i'-1}p_{i'-1}\right) \right. \nonumber \\
 & + & \left. \left( \Gamma _{i'}V_{i'}\alpha _{i'}Q_{i'}
-\Gamma _{i'-1}V_{i'-1}\alpha _{i'-1}Q_{i'-1}\right) \right] \nonumber \\
 & + & \frac{\alpha _{i}}{\overline{r}_{i}}\left[
\left( 1+\frac{e_{i}}{c^2}\right)
\left( 1+\frac{6V_{i}\left( p_{i}+Q_{i}\right) }{m_{i}c^2}\right)
\frac{Gm_{i}}{r_{i}}\right. \nonumber \\
 & + & \left. \frac{1}{\rho _{i}c^2}\left( u_{i}^{2}\left( 2p_{i}-Q_{i}\right)
-\frac{Gm_{i}}{r_{i}}\left( p_{i}+Q_{i}\right) \right) \right] da_{i} \nonumber \\
 & - & \alpha_{i'-1}\overline{\Gamma }_{i}S_{i}^{{\rm ext}}
   - \alpha_{i'-1}\overline{u}_{i} e_{i'-1}^{\rm ext}=0.
\mylabel{eq_fd_momentum}
\end{eqnarray}
We now use the sum of the internal energy equation and velocity times the momentum
equation in the Newtonian limit as a guideline for the discretization of the
total energy equation. In the Newtonian limit, and with the omission of external
sources, advection, and artificial viscosity we find
\begin{eqnarray}
F^{e}_{i'}+u_{i+1}F^{S}_{i+1} & = & \frac{e_{i'}da_{i'}-\overline{e}_{i'}\overline{da}_{i'}}{dt}+p_{i'}\left( w_{i+1}-w_{i}\right) \nonumber \\
 & + & u_{i+1}\frac{u_{i+1}da_{i+1}-\overline{u}_{i+1}\overline{da}_{i+1}}{dt}
\nonumber \\
 & + & w_{i+1}\left( p_{i'+1}-p_{i'}\right) +\frac{u_{i+1}}{\overline{r}_{i+1}}\frac{Gm_{i+1}}{r_{i+1}}da_{i+1}.
\end{eqnarray}
The expression \( w_{i+1}p_{i'} \) from the internal energy equation cancels
with its counterpart from the momentum equation and leaves the adiabatic work
\( w_{i+1}p_{i'+1}-w_{i}p_{i'} \) that is applied to the fluid element. Which
fluid element? We observe that the conservation of energy is expressed by relating
the internal energy enclosed by zone edges, and the kinetic and gravitational
energy enclosed by zone centers, to this adiabatic work. The work excerted on
such a ``distorted zone'' is given by the product of the volume changes \( w \)
at zone edges with the pressure \( p \) at zone centers. The position of these
terms perfectly fits the arrangement of the variables on the staggered grid
(and has to be remembered when one implements energy conservation checks).
This suggests the following discretization of the full energy equation
(\myref{eq_total_energy})
\begin{eqnarray}
\left( \tau 1_{i'}da_{i'}+\tau 2_{i+1}da_{i+1}+\tau 3_{i+1}da_{i+1}\right) \frac{1}{dt} &  &  \nonumber \\
-\left( \overline{\tau 1}_{i'}\overline{da}_{i'}+\overline{\tau 2}_{i+1}\overline{da}_{i+1}+\overline{\tau 3}_{i+1}\overline{da}_{i+1}\right) \frac{1}{dt} &  & \nonumber \\
+\tau 1_{i+1}^{{\rm adv}}+\tau 2_{i'+1}^{{\rm adv}}+\tau 3^{{\rm adv}}_{i'+1}-\tau 1^{{\rm adv}}_{i}-\tau 2_{i'}^{{\rm adv}}-\tau 3^{{\rm adv}}_{i'} &  & \nonumber \\
+w_{i+1}\alpha _{i'+1}\left[ \frac{1}{2}\left( p_{i'+1}+\overline{p}_{i'+1}\right) +Q_{i'+1}\right]  &  & \nonumber \\
-w_{i}\alpha _{i'}\left[ \frac{1}{2}\left( p_{i'}+\overline{p}_{i'}\right) +Q_{i'}\right] -\alpha_{i'}\overline{\Gamma }_{i+1}e_{i'}^{{\rm ext}}-\alpha_{i'}\overline{u}_{i+1}S_{i+1}^{{\rm ext}} & = & 0.
\mylabel{eq_fd_total_energy}
\end{eqnarray}
 It is problematic to apply time centering on the adaptive grid because the
quantities belong to different fluid elements across a time step. The finite
differencing is therefore kept first order backward Euler. Two exceptions, however,
improved the accuracy of the solution considerably while they were not found
to conflict with the grid motion. In the total energy equation, we time-average
the pressure as in \citep{Swesty_95} and in the momentum equation we
time-average the \( 1/r^{2} \) term in the gravitational force as in \citep{Yamada_97}.
We solve the total energy equation (\myref{eq_fd_total_energy}) together with
the momentum equation
(\myref{eq_fd_momentum}) or the internal energy equation
(\myref{eq_fd_internal_energy}) depending on the contribution of
internal and kinetic energy to the total energy. As discussed in the previous
section, the momentum equation and the internal energy equation are merged
with weights according to 
\begin{equation}
e_{i'}F^{S}_{i+1}-\frac{1}{2}S_{i+1}F^{e}_{i'}=0.
\mylabel{eq_fd_mix}
\end{equation}
 The evolution of the electron fraction is determined by \( Y^{{\rm ext}}_{e} \)
that describes weak interactions with neutrinos 
\begin{equation}
\frac{Y_{e,i'}da_{i'}-\overline{Y}_{e,i'}\overline{da}_{i'}}{dt}+Y_{e,i+1}^{{\rm adv}}-Y_{e,i}^{{\rm adv}}-\alpha_{i'}Y^{{\rm ext}}_{e,i'}=0.
\label{eq_fd_ye}
\end{equation}
The eight equations
(\myref{eq_fd_volume_gradient})-(\myref{eq_fd_continuity})
and (\myref{eq_fd_total_energy})-(\myref{eq_fd_ye})
are implicitly solved in one time step,
together with the adaptive grid equation (\myref{eq_adaptive_grid}).

\subsection{Boundary conditions}

For completeness, we add in this section details about the implementation
of the boundary conditions at the border of the computational domain.
The boundary conditions are partially modified for the successful run
of the individual test problems as described below.

We set the zone edge with \( i=1 \) at the center of spherical symmetry
and impose boundary conditions
\begin{eqnarray}
a_{1} - \overline{a}_{1} & = & 0
\label{eq_fd_boundary_a1} \\
r_{1} - \overline{r}_{1} & = & 0 \\
u_{1} - \overline{u}_{1} & = & 0 \\
m_{1} - \overline{m}_{1} & = & 0
\label{eq_fd_boundary_m}
\end{eqnarray}
for the variables that live on zone edges. The reason for rather
keeping them constant instead of explicitly setting them to zero
is that this formulation remains valid in simulations where the
inner boundary is off center, as e.g. in the shock tube calculations
presented in the next section. Special care has to be taken of the
advective fluxes at the zone center \( i=1+1/2 \). Although the
kinetic and potential energy, as well as the radial momentum,
vanish at the center, the content of the first half zone
with respect to these conserved quantities is not zero. It
is subject to advection if the zone center \( i=1+1/2 \) moves
inwards. In the following reasoning about the distribution of
these quantities around the symmetry center, we will neglect
the difference between the gravitational and rest mass, and
set \( \Gamma=1 \). We assume
that \( a/r^3 = A \) is constant in the neighborhood
of the symmetry center and estimate for the gravitational
energy in a sphere
\begin{equation}
\int_{0}^{a} \frac{a}{r} da =
\frac{3}{5} A^{\frac{1}{3}}a^{\frac{5}{3}}.
\end{equation}
For the evaluation of the kinetic energy and radial momentum,
we assume \( u/r = B \) constant and evaluate
\begin{eqnarray}
\int_{0}^{a} \frac{1}{2} u^2 da & = &
\frac{3}{10} B^2 A^{-\frac{2}{3}} a^{\frac{5}{3}} \nonumber \\
\int_{0}^{a} u da & = &
\frac{3}{4} B A^{-\frac{1}{3}} a^{\frac{4}{3}}.
\end{eqnarray}
The amount of the conserved quantity in the first half
zone is then found by integrating the conserved quantity
over the first one and a half zones from \( a_{1}=0 \) to
\( a_{2'} \) and subtracting the content that has already
been taken account of in the range \( da_{2} \). With an
equidistant zoning, \( a_{2'} = 3/2 da_{2} \), we find
the relations
\begin{eqnarray}
\int_{0}^{a_{2'}} \frac{a}{r} da  & = &
\frac{3}{5}\left( \frac{3}{2} \right)^{\frac{5}{3}} \frac{a_2}{r_2}da_2
\nonumber \\
\int_{0}^{a_{2'}} \frac{1}{2} u^2 da  & = &
\frac{3}{5}\left( \frac{3}{2} \right)^{\frac{5}{3}} \frac{1}{2} u_2^2 da_2
\nonumber \\
\int_{0}^{a_{2'}} u da  & = &
\frac{3}{4}\left( \frac{3}{2} \right)^{\frac{4}{3}} u_2 da_2
\end{eqnarray}
that express the content of the innermost one and a half zones in
terms of the content in the interval \( da_{2} \) that is well placed
on the staggered grid. We therefore define the fractions
\( f_{e} = 3/5(3/2)^{5/3}-1 \),
\( f_{S} = 3/4(3/2)^{4/3}-1 \), and
set for advection purposes a content
\begin{eqnarray}
\tau 2_{1} & = & f_{e} \tau 2_{2} \nonumber \\
\tau 3_{1} & = & f_{e} \tau 3_{2} \nonumber \\
S_{1} & = & f_{S} S_{2}
\end{eqnarray}
for the kinetic energy, gravitational energy, and the radial momentum in the
innermost half zone. These measures are not required in the simulations
of the shock tubes, where the inner boundary is off center and the physical
state constant in its neighborhood.

At the outer boundary, we impose the conditions
\begin{eqnarray}
\rho_{n'} - \overline{\rho}_{n'} & = & 0
\label{eq_fd_boundary_n} \\
T_{n'} - \overline{T}_{n'} & = & 0
\label{eq_fd_boundary_T} \\
Y_{e,n'} - \overline{Y}_{e,n'} & = & 0
\label{eq_fd_boundary_Ye} \\
\alpha_{n'} - \left( 1 - \frac{2m_{n'}}{r_{n'}} \right) \Gamma_{n'}^{-1}
& = & 0
\label{eq_fd_boundary_alpha} \\
a_{n} - \overline{a}_{n} & = & 0.
\label{eq_fd_boundary_a2}
\end{eqnarray}
We determine the quantities \( m \), \( r \), and \( u \) on the last
zone center \( n' \) from the last zone edge \( n \) with the
equations
(\myref{eq_fd_volume_gradient}), (\myref{eq_fd_mass_gradient}),
and \( w_{n'} - w_{n} = 0 \). Equations
(\myref{eq_fd_boundary_n})-(\myref{eq_fd_boundary_Ye}) implement
a constant surface pressure boundary condition by the equation
of state. Equation (\myref{eq_fd_boundary_alpha}) matches the
lapse function to the exterior Schwarzschild metric. Note that
in equations (\myref{eq_fd_boundary_a1})-(\myref{eq_fd_boundary_m})
and (\myref{eq_fd_boundary_n})-(\myref{eq_fd_boundary_a2}) every
independent variable fulfills one boundary condition, except for
the rest mass \( a \). The computational domain is tied to the
desired mass range by two boundary conditions that compensate for
the missing boundary condition for \( n/R \) in the adaptive grid
equation (\myref{eq_adaptive_grid}). The constant pressure boundary
condition is not very practical for the homologous collapse test
because the surface tends to drift outwards during the
time evolution and destroys the homologous velocity profile.
We therefore evolve the surface pressure proportional to the
change in the central pressure and replace equation
(\myref{eq_fd_boundary_n}) and (\myref{eq_fd_boundary_T}) by
\begin{eqnarray}
p_{n'} - \frac{p_{1'}}{\overline{p}_{1'}}p_{n'}^{\rm init} & = & 0 \\
T_{n'}/\rho_{n'}^{\gamma -1} - \overline{T}_{n'}/\overline{\rho}_{n'}^{\gamma -1} & = & 0.
\end{eqnarray}
The latter condition imposes constant entropy at the surface.


\section{Supernova-Related Standard Test Problems}
\label{section_tests}

In this section we demonstrate
the capability of our hydrodynamics code to accurately reproduce important
features in stellar core collapse and supernova explosions.
We present the standard test calculations for supernova codes
\citep{Swesty_95,Yamada_97} that
complement the application of AGILE in \citep{Mezzacappa_et_al_01}
and \citep{Liebendoerfer_et_al_01}.
All examples are
based on a polytropic equation of state and, unless marked otherwise,
simulated with the same general relativistic code version.

\subsection{Sedov point-blast explosion}

A nonrelativistic point-blast explosion in spherical symmetry was
analytically analyzed by \citet{Taylor_50} and
\citet{Sedov_59} (we followed \citet{Landau_Lifschitz_91}
for the reproduction of the self-similar analytical solution to
this problem). An amount of energy \( E_{tot } \) is deposited in a
uniform gas with a density \( \rho _{0} \) and internal energy that is
negligible with respect to \( E_{tot} \). We start with an
initial specific energy \( e_{0}=10^{-3} \) at a density \( \rho _{0}=1
\) in cgs units. The pressure is given by the ideal gas equation of
state \( p=(\gamma -1)\rho e \) with \( \gamma =5/3 \). The
computational domain has a radius of \( r=1 \) around the center where
the energy \( E_{tot}=1 \) is deposited. The initial state is prepared
by starting with \( 100 \) equidistant grid points. The explosion
energy \( E_{tot} \) is placed into the innermost zones according to an
exponentially decaying spatial distribution on a time scale
between the dynamical time scale and the much shorter time scale
of the grid adaption. By this measure, the adaptive grid can move
inwards and resolve the exponential shape of the deposited energy. If
the amount of \( E_{tot} \) is resolved, the time scale
increases to the dynamical time scale and the shock wave starts to move
radially outwards.
\begin{figure}
  \plotone{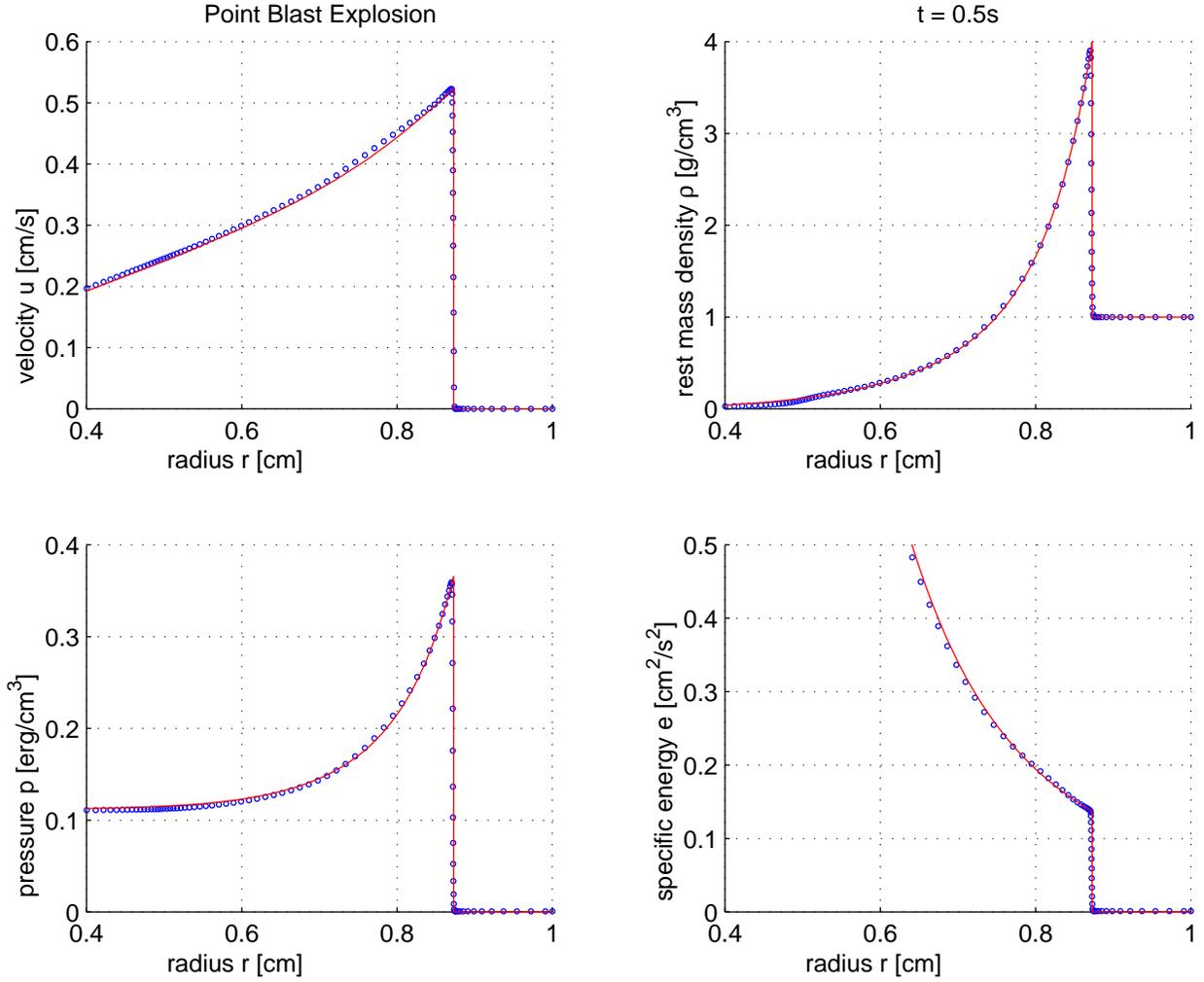}
  \caption{This figure shows the Sedov point blast explosion, the
    spherically symmetric explosion after the deposition of a large energy
    amount in the symmetry center at \( r=0 \). The dots indicate the location
    of the grid points in the numerical solution. They are compared to
    the solid line, which represents the analytical solution.}
  \mylabel{f3.ps}
\end{figure}
The numerical solution is compared to the analytical solution
at time \( t=0.5 \) in Fig. (\myref{f3.ps}).
We find the correct shock speed and emphasize the resolution
achieved with the adaptive grid. The deviations in the inner part
of the sphere stem from the nonideal initial conditions.

\subsection{Nonrelativistic and relativistic shock tube}

Probabely the most popular test is the calculation of a shock tube. The
initial conditions for a nonrelativistic simulation were proposed by
\citet{Sod_78} and a
special relativistic version was calculated by
\citet{Centrella_Wilson_84} and \citet{Marti_Mueller_94}.
Sod's shock tube evolves the decay
of a discontinuity in density \( \rho  \) and specific energy \( e \)
in a closed box. The box is filled with an ideal gas whose pressure is
given by a polytropic equation of state with adiabatic index \( \gamma
=7/5 \). In our spherical code we calculate the shock tube in a
spherical shell with thickness $d=4$ cm at a large radius of $r_{0}=10000$ cm
in order to approximate slab symmetry. In the initial state, the
discontinuity is placed at \( r_{0} \). The left hand side state is \(
\left\{ u,\rho ,e,p\right\} _{L}=\left\{ 0,1,2.5,1\right\}  \) and the
right hand side state is
\( \left\{ u,\rho ,e,p\right\} _{R}= \left\{0,0.125,2,0.1\right\}  \)
in cgs units. The discontinuity in the initial profile of the
variables \( y \) is smoothed around \( r_{0} \) by a hyperbolic tangent
function
\begin{equation}
y(r)=y(<r_{0})+\frac{1}{2}\left( y(>r_{0})-y(<r_{0})\right) \left(
1+\tanh \left ( g\left( r-r_{0}\right) \right) \right) \end{equation}
whith a profile slope of \( g=150 \) at the edge \( r_{0} \).
\begin{figure}
  \plotone{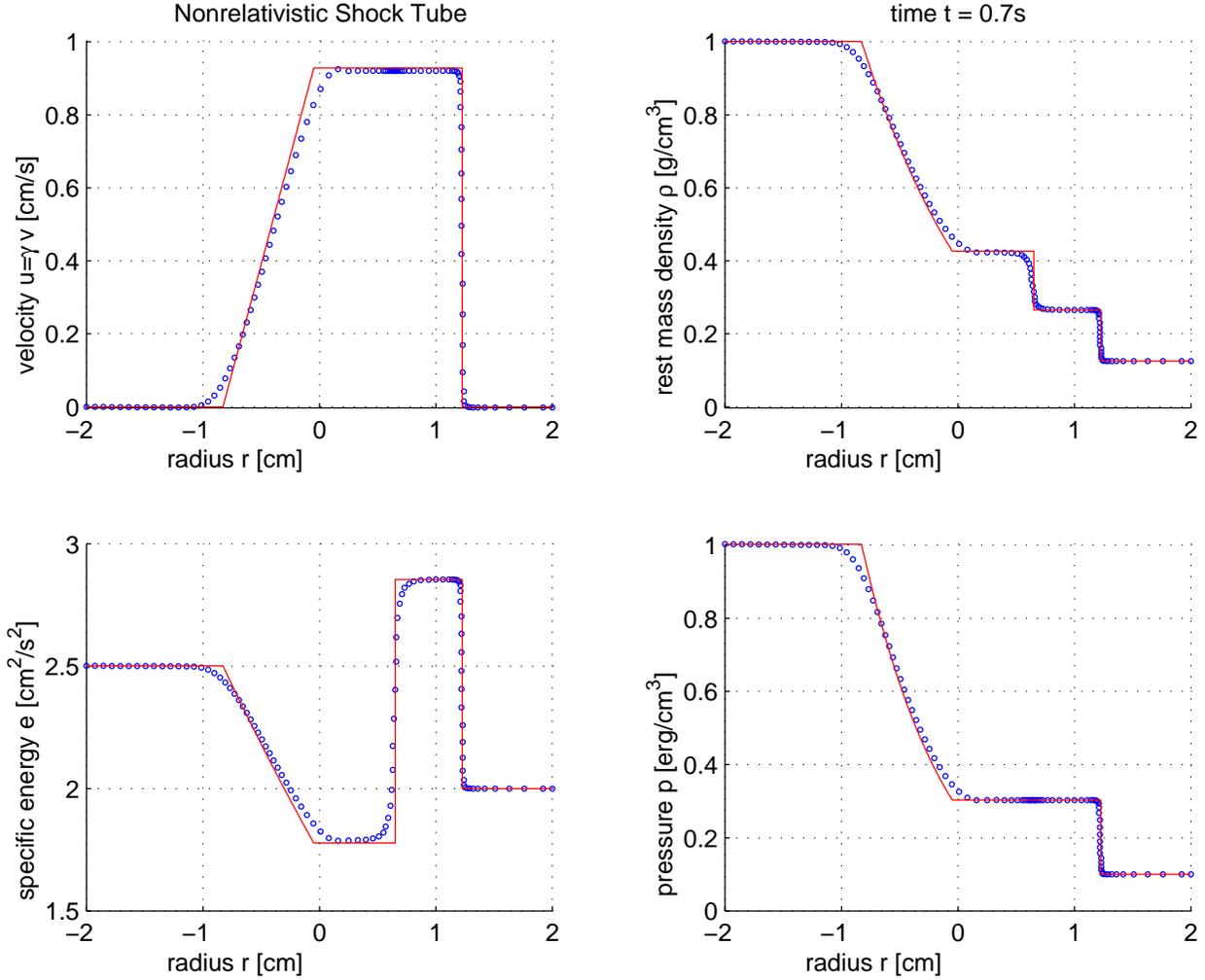}
  \caption{This figure shows a shocktube calculation with initial
    data leading to nonrelativistic velocities. The dots indicate
    the location of the grid points in the numerical solution. The
    solid line represents the analytical solution. In a time dependent
    visualization of this graph one would observe the grid points moving
    in accordance with the shock front. It is mainly the space coordinate
    of a grid point in the shock front that changes value, and not so
    much the represented physical state.}
  \mylabel{f4.ps}
\end{figure}
The comparison of the numerical and the
analytical results is shown in Fig. (\myref{f4.ps}).  As expected
from the conservative formulation, the shock speed exactly matches the
analytical solution and the jump conditions are fulfilled.
Somewhat smoother transitions in the numerical
calculation around the contact discontinuity and the rarefaction wave
are on the one hand due to the smoothened initial state and on the other
hand to the diffusion according to equation (\myref{eq_advective_diffusivity})
that is given by the advection scheme.

In order to test also the relativistic regime, we set up initial
conditions for a relativistic shock. The initial state is \( \left\{
u,\rho ,e,p\right\} _{L}=\left\{ 0,1,2.5\times 10^{22},10^{22}\right\}
\) for \( r<r_{0} \) and \( \left\{ u,\rho ,e,p\right\} _{R}=\left\{
0,0.125,2 \times 10^{22},10^{21}\right\}  \) for \( r>r_{0} \). The
velocity \( u \) reaches about half of the velocity of light. This is
much more than expected in any supernova calculation and enough to
distinguish relativistic effects in the result. In the reproduction
of the exact solution, we followed the approach of \citet{Centrella_Wilson_84}
and replaced their numerical integration of equation (C13) with
an analytical solution\footnote{
To ease the search for an analytical solution to the equivalent
of equation (C13) in \citep{Centrella_Wilson_84},
we introduce the abbreviations \( y=\Gamma E/D \) and
\( x=(1-V^2)/(V-\xi) \), where \( \Gamma \), \( E \), \( D \), \( V \),
and \( \xi \) have been defined as described in 
\citep{Centrella_Wilson_84}. They are not to be confused with the
notation in this paper. Equation (C12) can then concisely be written as
\begin{equation}
0 = ( x+V )^2 - \frac{1+y}{y(\Gamma -1)}.
\end{equation}
Instead of \( dD/dV \), leading to equation (C13), we rather evaluate
\begin{equation}
\frac{dy}{dV} = -\frac{y(\Gamma -1)}{1-V^2} (x+V) = -\sqrt{\Gamma -1}
\frac{\sqrt{y(y+1)}}{1-V^2}.
\end{equation}
After the substitution of \( z \) for the variable \( y \) according
to the relation \( y=1/2(cosh(z) - 1) \), we find with
\( dy=\sqrt{y(y+1)}dz \) the ODE
\begin{equation}
\frac{dz}{dV} = \frac{\sqrt{\Gamma -1}}{1 - V^2}
\end{equation}
that has the analytical solution
\( V = \tanh \left(\frac{z}{\sqrt{\Gamma -1}} \right) \).
}.
Analytical solutions to the special relativistic shocktube
problem in an inertial coordinate frame were also derived
by \citet{Marti_Mueller_94}. Note, however, that the comoving 
coordinates used in our code do not limit to a global
inertial frame in the special relativistic limit. While the
areal radius, \( r \), the enclosed rest mass, \( a \), and
the velocity, \( u \), share their definition in both,
the comoving Misner-Sharp coordinates and the inertial
Schwarzschild coordinates, the definition of coordinate time 
is not the same in these coordinate choices. The correspondence
is straightforward for the region at rest in the inertial frame and
the region containing the matter between the contact discontinuity
and the shock front because it is instantaneously accelerated
to constant speed as the shock passes through. The fluid elements
in the rarefaction wave have a more complicated
velocity history and show an individual time lapse with respect
to the inertial frame.
\begin{figure}
  \plotone{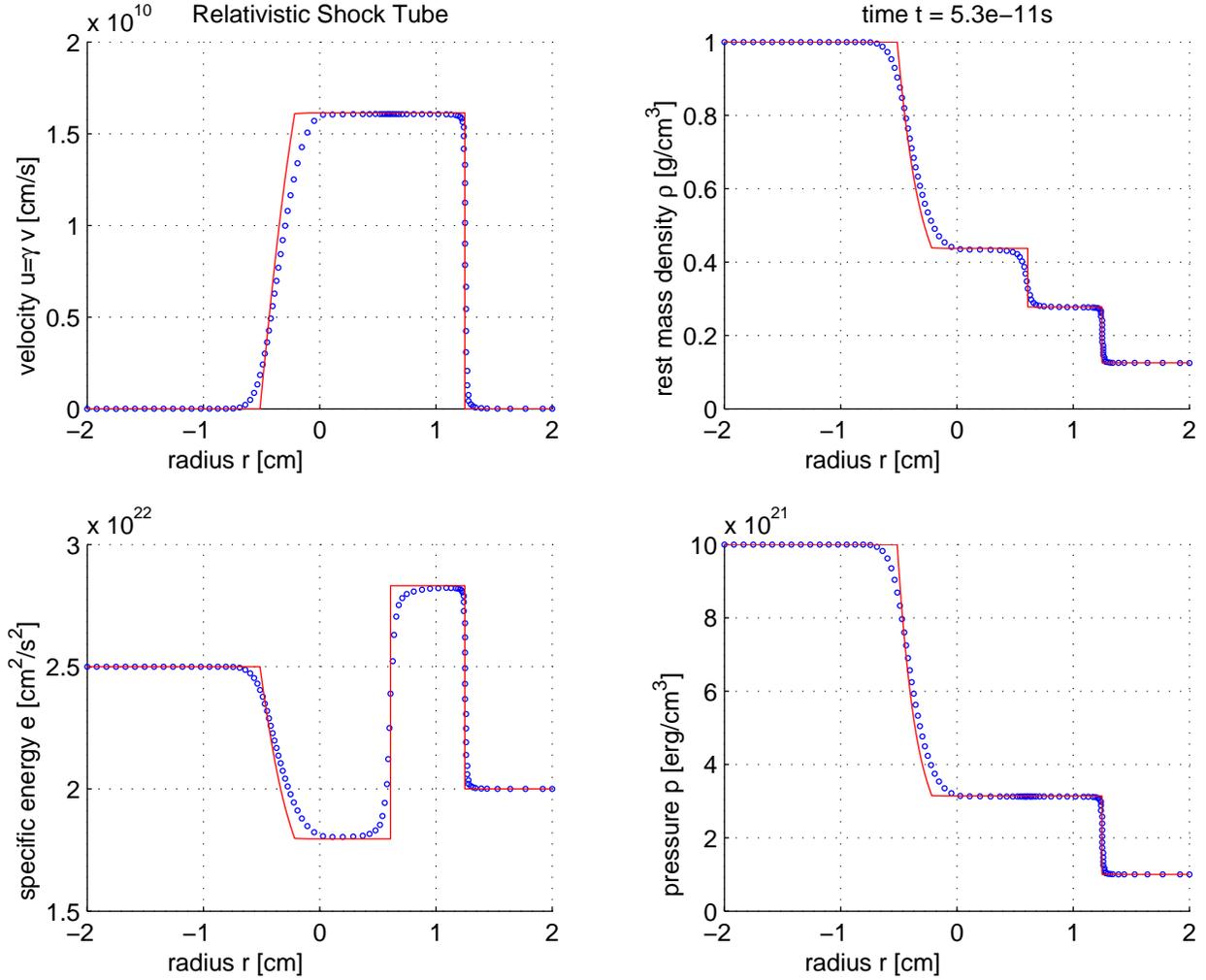}
  \caption{This figure shows a shocktube calculation that
    leads to relativistic velocities. The physical velocity
    \( v=u/\gamma \) is large enough to distinguish relativistic
    effects. The dots indicate
    the location of the grid points in the numerical solution. The
    solid line represents the analytical solution. Note that
    the time measured in our comoving coordinates differs from
    the time measured in an inertial frame.}
  \mylabel{f5.ps}
\end{figure}
In Fig. (\myref{f5.ps}) we compare the numerical solution
to the analytical solution.
This test demonstrates the accurate
implementation of the special relativistic conservation laws that
lead to the correct relativistic shock jump conditions.

\subsection{Oppenheimer-Snyder dust collapse}

The collapse of a uniform and pressureless dust cloud was first
calculated in general relativity by
\citet{Oppenheimer_Snyder_39}. Although most of the relativistic terms
vanish for negligible pressure, this test problem is often recommended
for general relativistic hydrocodes. It mainly tests the implementation
of the momentum equation, because each fluid element behaves like a free
particle in the self-gravitating pressureless gas. Initial configuration
is a dust cloud of two solar masses with a density of \( \rho =10^{8}gcm^{-3}
\). In order to assure a negligible pressure, we set the initial temperature
in the dust cloud to \( T=10^{-5}MeV \). We had to switch off the adaptive
grid for this run because spurious variations in the temperature profile
lead to grid motions with advection inaccuracies that showed a positive
feedback on the temperature variations.
\begin{figure}
  \plotone{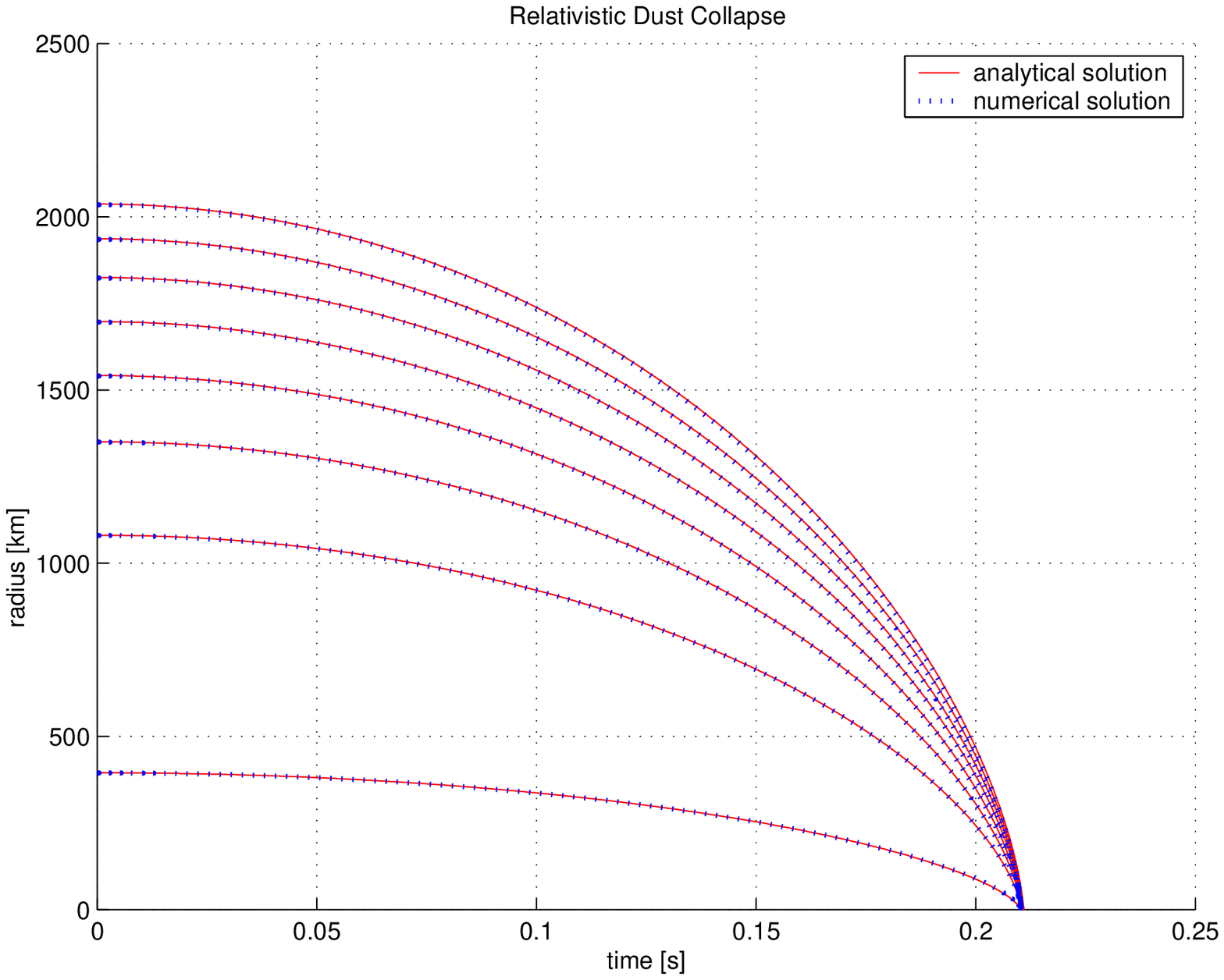}
  \caption{This figure shows the collapse of a homologous dust cloud in
    comoving coordinates. The numerical solution (dots) can hardly be
    distinguished from the analytical solution (solid lines) at early
    times. The lines represent mass traces at enclosed baryon masses
    of $[0.013,0.266,0.519,0.772,1.03,1.28,1.53,1.78]$ solar masses.}
  \mylabel{f6.ps}
\end{figure}
The result presented in Fig. (\myref{f6.ps}) demonstrates that
the match of the lapse function to the Schwarzschild metric outside
of the cloud and that the discretization of the momentum equation
are fine.

\subsection{Homologous collapse}

Here we present the homologous collapse that was first analyzed
by \citet{Goldreich_Weber_80}.
Again we use a polytropic equation of state with
\( p=\rho T/m_{b} \) and \( e=(\gamma -1)^{-1}T/m_{b} \), where \( T \)
is the temperature and \( \gamma =4/3 \) the adiabatic index.
With this equation of state, \( p = (\gamma -1)\rho e \), one easily
derives the time derivative
\begin{equation}
\frac{d}{dt} \left( \frac{p}{\rho^{\gamma}} \right)
= \frac{\gamma -1}{\rho^{\gamma -1}} \left[
\frac{de}{dt} + p \frac{d}{dt}
\left( \frac{1}{\rho} \right) \right]=0.
\end{equation}
Thus, the quantity \( p/\rho^{\gamma} \) is expected to remain
constant during the dynamical evolution. Our finite difference
representation is chosen to conserve total energy to machine
precision. The check for constant entropy in an adiabatic evolution
therefore provides important information about the accuracy of
the evolution of the internal energy. We run this test
with the same artificial viscosity as in the realistic simultions
in order to check its (non-)influence during the homologous 
compression in the collapse phase.
Following \citet{Yamada_97} we start with a hydrostatic
isentropic state with central density \( \rho _{c}=10^{8}\) g cm$^{-3}$
and temperature \( T_{c}=0.2\) MeV.  This configuration is uniformly
cooled until we achieve marginal stability. We then reduce the pressure
further by a total of $3$\% \citep{Goldreich_Weber_80},
distributed over \( 300 \) time steps. This leads to
a smooth transition into collapse without initiating pressure
waves that disturb the infall of the surface and lead to
dissipation of kinetic energy in small shocks. We update the
surface pressure proportionally to the evolution of the central
pressure in order to prevent the surface from drifting out of
the homologous motion owing to the lack of an outer part of the
star. The test is carried out in the Newtonian limit,
the fully relativistic run shows a less homologous velocity
profile, but similar deviations in the enetropy.
\begin{figure}
  \plotone{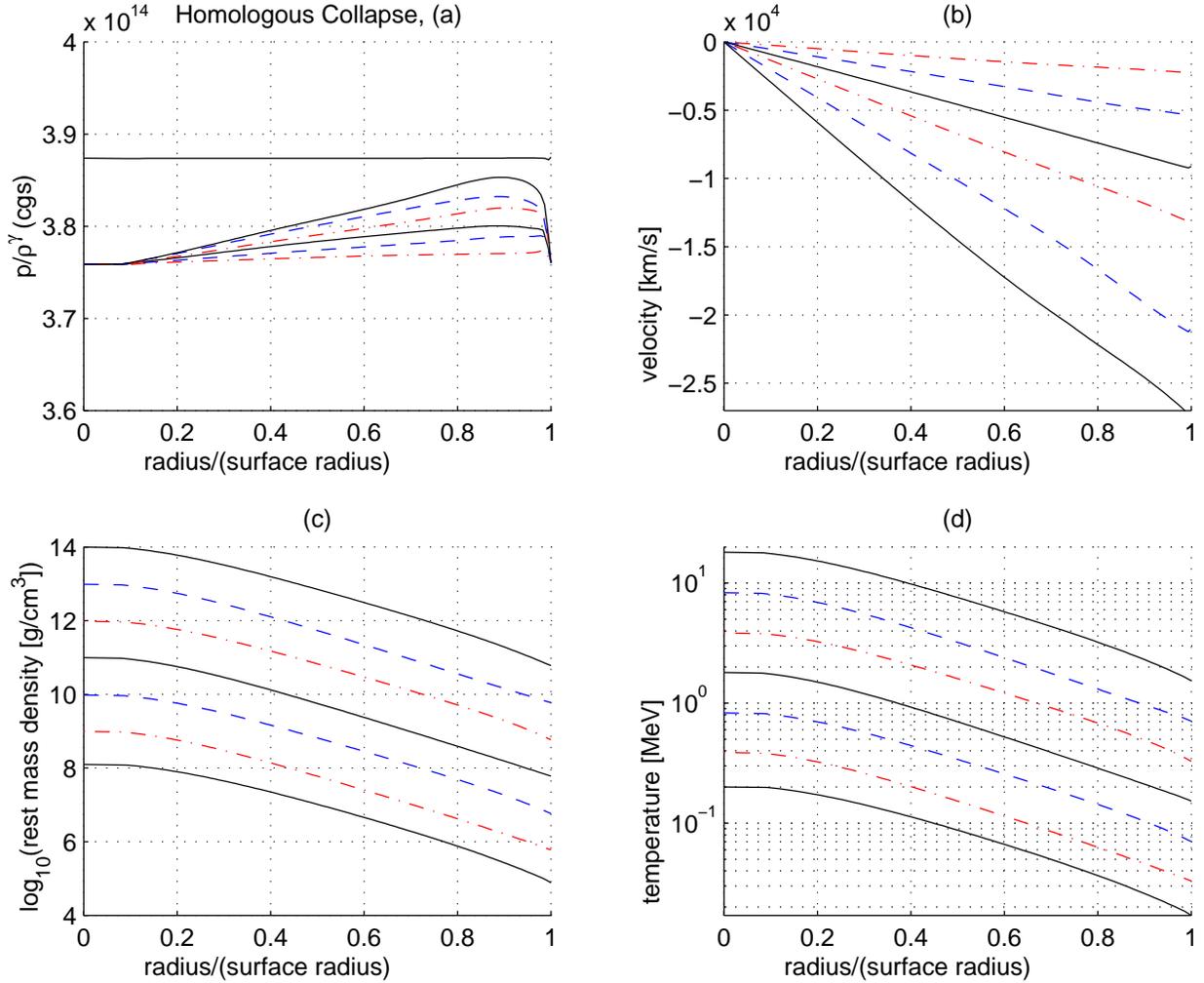}
  \caption{The homologous collapse of an isentropic polytrope is
    shown. Snapshots are taken when the central density reaches
    $[10^8,10^{9},10^{10},10^{11},10^{12},10^{13},10^{14}]$ g cm$^{-3}$. Graph
    (a) shows the entropy evolution. The uppermost solid line
    belongs to the initial state. The lowermost dash-dotted line
    belongs to a central density of $10^9$ g cm$^{-3}$ and clearly
    shows the $3$\% pressure reduction applied to initiate the
    collapse. The entropy deviation grows with increasing
    compactness. Graph (b) shows the velocity profiles. Graph
    (c) and (d) visualize the selfsimilar evolution of density
    and temperature.}
  \mylabel{f7.ps}
\end{figure}
The homologous collapse of an isentropic polytrope is
shown in Fig. (\myref{f7.ps}). Snapshots are taken for
every decade in the central density. The radius of the profiles
is normalized to the surface radius. Graph (a) shows the entropy
evolution. The uppermost solid line belongs to the initial state.
The lowermost dash-dotted line belongs to a central density of
$10^9$ g cm$^{-3}$ and clearly monitors the $3$\% pressure reduction
applied to initiate the collapse. The entropy deviation grows
with increasing compactness, but does not exceed the $3$\% level
until nuclear density is achieved in the center of the star.
Graph (b) shows the nicely homologous velocity profiles with
the velocity being proportional to the radius. Graph (c) and
(d) visualize the selfsimilar evolution of density and temperature.

\subsection{Comparison to PPM}
\label{section_PPM}

Finally, we compare the shock tube result calculated with AGILE to
the shock tube data provided by Raph Hix and Alan Calder from the
EVH-1 implementation of PPM. Piecewise Parabolic Methods
became a standard in hydrodynamics codes while conventional finite
difference schemes gathered some patina, illustrated by their
dependence on the unspeakable artificial viscosity.
\begin{figure}
  \plotone{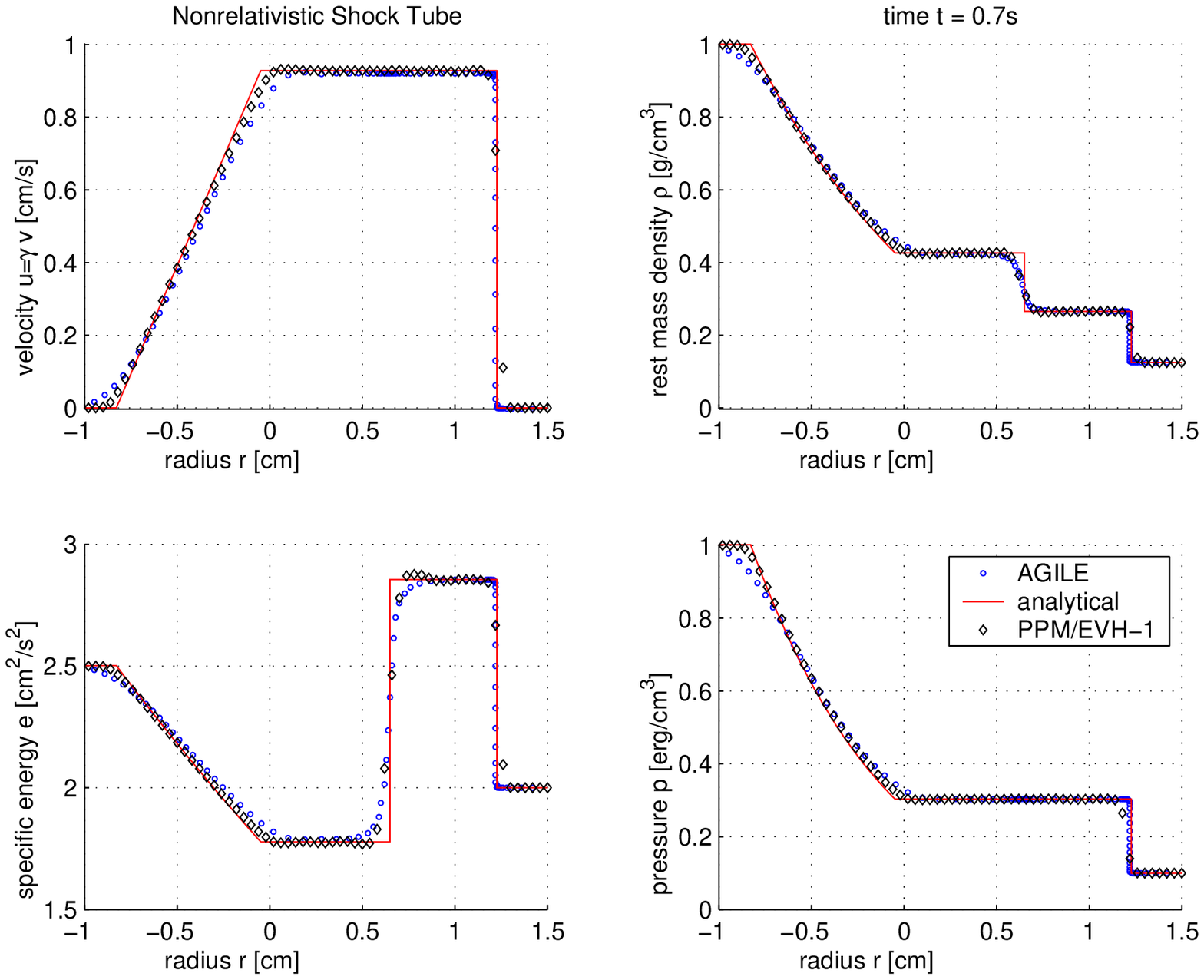}
  \caption{A nonrelativistic shocktube calculation comparing the
    results of AGILE and EVH-1. Both codes use \( 100 \) grid points.}
  \label{f8.ps}
\end{figure}
In Fig. (\myref{f8.ps}), a comparison between the two methods
is shown and related to the analytical solution. The solution of
EVH-1 was calculated on \( 100 \) equidistant grid points while AGILE used
the same number of adaptive grid points. The reconstruction routines
in EVH-1 do not implement a contact discontinuity steepening. A flattening
parameter \( f \le 1 \) is introduced to provide additional dissipation
in the shock if a relative pressure jump exceeds \( \varepsilon=0.33 \)
as defined by \citet{Colella_Woodward_84} in Eq. (A.1). Following their
prescription given in Eq. (A.2), this flattening parameter has been set
to
\begin{equation}
\tilde{f}_{j} = 1 - \max \left( 0, \left( \frac{p_{j+1}-p_{j-1}}
{p_{j+2}-p_{j-2}} - \omega^{(1)} \right) \omega^{(2)} \right)
\end{equation}
with \( \omega^{(1)}=0.75 \) and \( \omega^{(2)}=5 \).
In AGILE, we have set a relatively
small artificial viscosity with \( \Delta l=2\times 10^{-3} \) cm in order
to bring the misunderstanding about accuracy and artificial
viscosity into prominence.
The adaptive grid technique renders an unrivalled resolution in
the shock front. A simulation on an equidistant grid would require
\( 32000 \) grid points to achieve a similar resolution at the shock.
Moreover, the shock
position and plateaus in the physical state are accurate.
With a conservative formulation of the equations, the
implicit algorithm solves numerically for the same
jump conditions at the shock front as are solved analytically
in a Riemann solver. The artificial viscosity only affects the
shock width and provides the means for energy dissipation in
the shock that is always present in a realistic fluid, even if
the shock spreads over a narrower spatial range. The weakness
of the adaptive grid, however, is the rarefaction wave. The
diffusivity induced by {\em advection} on the dynamical adaptive
grid as discussed in section \myref{section_diffusion} is responsible
for the more indistinct rarefaction wave and contact discontinuity.
According to equation (\ref{eq_advective_diffusivity}), the artificial
diffusion is large at increased grid spacing and high grid velocities.
As expected, the PPM result shows less difficulty in the
rarefaction wave, owing to its more
uniform grid point distribution and smaller diffusivity.
The run with EVH-1 required \( 40 \) time steps with \( 4\times 10^{-4} \)
CPU seconds per explicit time step on an IBM RS/6000 SP, the run with AGILE required
\( 429 \) time steps with \( 0.0713 \) CPU seconds per implicit time step.
\( 295 \) time steps were used until physical time \( 0.35 \) s and
\( 134 \) time steps were needed from \( 0.35 \) to \( 0.7 \) s. The efficiency
increases in the second part of the simulation because, as illustrated
in Fig. (\myref{f2a.ps}), the adaptive
grid can take advantage of the self-similarity of the flow. About \( 30 \% \)
of the computational effort is spent with the inversion of the Jacobian.
There are many ways to compare the performance depending on the importance
of specific features. If one would request the same maximum resolution,
the number of time steps required in EVH-1 would dramatically scale up by
a factor of \( 320 \) due to the Courant restriction on the time step. 
It is also important to note that the CPU time per time step is
not relevant for the realistic applications we have in view.
The computational effort is dominated
by the radiation transport or the nuclear reaction network. Important
is the overall number of zones and the ability to allocate them
in regions with interesting physics. Moreover, the total computational
effort will be proportional to the number of time steps. In the shock tube
calculation, the Courant condition of explicit schemes is only challenged
by resolution. In the supernova application, the restriction is much more
severe in the compact proto-neutron star, where the sound speed quickly
reaches a third of the velocity of light.


\section{Conclusion}

We report on AGILE, a first order implicit solver for
stiff algebro-differential equations.
AGILE builds the Jacobian by automatized numerical finite
differences. It is therefore flexible with respect to
changes and extensions in the physical equations and
robust against programming errors in the now obsolete
implementation of the Jacobian. As discussed in the
introduction, implicit finite differencing
is especially useful in astrophysical applications where
the characteristic time scale of physical processes varies by orders of
magnitude.

In spherical symmetry, we have written the Einstein
equations in comoving coordinates and conservative form.
These equations are implemented in a module for AGILE to
obtain a spherically symmetric general relativistic
hydrodynamics code. We recognize that a dynamical adaptive
grid is nothing else than a specific choice of shift
vectors in the view of a general relativistic 3+1 decomposition.
The adaptive grid technique of 
\citet{Winkler_Norman_Mihalas_84} and
\citet{Dorfi_Drury_87}, that was developed for Newtonian
hydrodynamics, has been extended to relativistic
space-time and used as a recipe for a stable
implementation of shift vectors that focus the numerical
work to physically important regions in the simulation.

With an application in stellar core collapse and
supernova simulations in view, special care of energy
conservation was taken. We present the
detailed wrinkles in our finite difference representation
of the physical equations. Our guidelines were (i) a
consistent discretization of interdependent constituents
to achieve an exact implementation of physical conservation
laws, (ii) avoidance of cancellation errors by the independent
evolution of small quantities, (iii) the absorption of
numerical defects into the physical concept of viscosity
for an overall consistency. Thus, the discretization of the
total energy equation is matched to the finite difference
representation of the momentum and internal energy equation
and leads to energy conservation at machine precision.
We quantify the diffusion introduced by advection in the
finite difference representation of the momentum equation.
Energy conservation requires a compensating pressure term
in the energy equation that we consistently absorbe into the
concept of artificial viscosity. Our artificial viscosity
is based on the tensor viscosity approach of
\citet{Tscharnuter_Winkler_79} that was extended to
general relativity by direct inclusion of the viscous stress-energy
tensor into the derivation of the Einstein equations
\citep{Liebendoerfer_Mezzacappa_Thielemann_01}.

We achieve accurate results in the standard test problems for
the supernova application, i.e., Sedov point blast explosion,
nonrelativistic shock tube, relativistic shock tube,
Oppenheimer-Snyder dust collapse, and in the homologous collapse.
We also compare our shock tube solution to a simulation with
the Piecewise Parabolic Method and demonstrate
a superior resolution of the shock with the adaptive grid. The
overall reduction of computational zones, without compromise in
the resolution of physically important regions, is especially helpful
if the hydrodynamics code is coupled to computationally more expensive
processes, as e.g. Boltzmann radiation transport or a nuclear reaction
network. The comparison with PPM confirms that the artificial
viscosity in the adaptive grid technique does not affect
the accuracy of the shock speed and the fulfillment of the correct
jump conditions. However, with the low order advection scheme chosen
in this work, considerable diffusion happens in the rarefaction wave
where the adaptive grid does not opt for a high resolution.

A previous version of AGILE has been used to simulate the hydrodynamical
part in X-ray burst simulations on an accreting neutron star
\citep{Rembges_99}. An AGILE module implementing
general relativistic hydrodynamics in polar slicing and radial gauge
has been tested and extended by \citet{Mueller_00}
to study the effect of dark matter in the formation of supermassive
black holes in the early universe.
A Newtonian module in Lagrangian coordinates has been used in
stellar core collapse and postbounce evolution simulations with complete
O(v/c) Boltzmann neutrino transport
\citep{Mezzacappa_et_al_01}. The module described in this paper,
in combination with general relativistic Boltzmann neutrino transport,
lead to the general relativistic simulation of stellar core
collapse and postbounce evolution \citep{Liebendoerfer_et_al_01}.


\acknowledgments
We thank Ewald M\"uller and Ernst Dorfi for helpful
guidance in the initial phase of the project and Anthony Mezzacappa
for motivating discussions and scientific support. We are greatly indebted
to Felix Rembges, Horst M\"uller, and Jacob Fisker for working with AGILE
and suggesting improvements. We are grateful for the data
provided to us by Raph Hix, Alan Calder, and Steve Bruenn for comparisons.
We enjoyed fruitful discussions with Wolfgang Hillebrandt,
Bronson Messer, and Christian Cardall.
M.L. is supported by the National Science Foundation under contract 
AST-9877130 and, formerly, was supported by the Swiss National 
Science Foundation under contract 20-47252.96 and 20-53798.98. 
S.R. acknowledges the support of a PDRA funded by PPARC.
F.-K.T. is supported in part by the Swiss National Science Foundation 
under contract 20-61822.00 and in part by the Oak Ridge National Laboratory,
managed by UT-Batelle, LLC, for the U.S. Department of Energy under
Contract DE-AC05-00OR22725.


\end{document}